\newif\ifTwoColumn
\newif\ifSUBMIT
\newif\ifCOMMENTS
\newif\ifFIGs
\newif\ifFIGoneColumn
\let\ifTwoColumn\iftrue
\let\ifSUBMIT\iftrue
\let\ifCOMMENTS\iffalse
\let\ifFIGs\iftrue
\let\ifFIGoneColumn\iftrue
\documentclass[10pt,a4paper]{iopart}
\usepackage{graphicx,cite}
\usepackage{bm,bbm}
\usepackage{iopams}
\usepackage[T1]{fontenc}
\usepackage[utf8]{inputenc}
\usepackage{graphicx}
\usepackage[english]{babel}
\usepackage{amsmath,amssymb,amsfonts}
\usepackage{ae}
\usepackage{xcolor}
\usepackage{bm,bbm,bbold}
\usepackage{enumerate}
\usepackage{math4phy}
\usepackage{booktabs}
\usepackage{listings}

\ifSUBMIT
  \ifCOMMENTS
    \usepackage{color}    
    \usepackage[normalem]{ulem}
    \def\EDITS#1{{\color{green}#1}}
    \def\STRIKE#1{{\color{red}\sout{#1}}}
    
    \def\NSTRIKE#1{{\color{blue}\sout{#1}}}
  \else
    \def\EDITS#1{#1}
    \def\STRIKE#1{}
    
    \def\NSTRIKE#1{}
  \fi
\else
  \usepackage{color}    
  \usepackage[normalem]{ulem}
 \definecolor{mygreen}{RGB}{0,180,0}    
 \definecolor{myred}{RGB}{200,0,0}    
  \def\EDITS#1{{\color{mygreen}#1}}
  \def\STRIKE#1{{\color{myred}\sout{#1}}}

  \def\NSTRIKE#1{{\color{blue}\sout{#1}}}
\fi
\usepackage{amsthm}
\newtheorem{thm}{Theorem}
\newtheorem*{mydef}{Definition}
\begin{document}

\renewcommand{\vec}[1]{\boldsymbol{#1}}
\renewcommand{\aa}{\vec{a}} 
\newcommand{\bb}{\vec{b}} \newcommand{\BB}{\vec{B}}
\newcommand{\cc}{\vec{c}} \newcommand{\CC}{\vec{C}}

\ifdefined\ee
  \renewcommand{\ee}{\vec{e}} \newcommand{\EE}{\vec{E}}
\else
  \newcommand{\ee}{\vec{e}} \newcommand{\EE}{\vec{E}}
\fi

\newcommand{\ff}{\vec{f}} \newcommand{\FF}{\vec{F}}
\newcommand{\hh}{\vec{h}} \newcommand{\HH}{\vec{H}}
\newcommand{\ii}{\vec{i}} \newcommand{\II}{\vec{I}}
\newcommand{\jj}{\vec{j}} \newcommand{\JJ}{\vec{J}}
\newcommand{\kk}{\vec{k}} \newcommand{\KK}{\vec{K}}
\newcommand{\mm}{{\vec{m}}} \newcommand{\MM}{\vec{M}}
\newcommand{\nn}{\vec{n}} \newcommand{\NN}{\vec{N}}
\newcommand{\oo}{\vec{o}} \newcommand{\OO}{\vec{O}}
\newcommand{\pp}{\vec{p}} \newcommand{\PP}{\vec{P}}
\newcommand{\qq}{\vec{q}} \newcommand{\QQ}{\vec{Q}}
\newcommand{\rr}{\vec{r}} \newcommand{\RR}{\vec{R}}
\newcommand{\vu}{\vec{u}} \newcommand{\UU}{\vec{U}}
\newcommand{\vv}{\vec{v}} \newcommand{\VV}{\vec{V}}
\newcommand{\xx}{\vec{x}} \newcommand{\XX}{\vec{X}}
\newcommand{\yy}{\vec{y}} \newcommand{\YY}{\vec{Y}}
\newcommand{\zz}{\vec{z}} \newcommand{\ZZ}{\vec{Z}}

\newcommand{\ua}{\mathrm{a}} \newcommand{\uA}{\mathrm{A}}
\newcommand{\ub}{\mathrm{b}} \newcommand{\uB}{\mathrm{B}}
\newcommand{\uc}{\mathrm{c}} \newcommand{\uC}{\mathrm{C}}
\newcommand{\ud}{\mathrm{d}} \newcommand{\uD}{\mathrm{D}}
\newcommand{\ue}{\mathrm{e}} \newcommand{\uE}{\mathrm{E}}
\newcommand{\uf}{\mathrm{f}} \newcommand{\uF}{\mathrm{F}}
\newcommand{\ug}{\mathrm{g}} \newcommand{\uG}{\mathrm{G}}
\newcommand{\uh}{\mathrm{h}} \newcommand{\uH}{\mathrm{H}}
\newcommand{\ui}{\mathrm{i}} \newcommand{\uI}{\mathrm{I}}
\newcommand{\uj}{\mathrm{j}} \newcommand{\uJ}{\mathrm{J}}
\newcommand{\uk}{\mathrm{k}} \newcommand{\uK}{\mathrm{K}}
\newcommand{\ul}{\mathrm{l}} \newcommand{\uL}{\mathrm{L}}
\newcommand{\um}{\mathrm{m}} \newcommand{\uM}{\mathrm{M}}
\newcommand{\un}{\mathrm{n}} \newcommand{\uN}{\mathrm{N}}
\newcommand{\uo}{\mathrm{o}} \newcommand{\uO}{\mathrm{O}}
\newcommand{\up}{\mathrm{p}} \newcommand{\uP}{\mathrm{P}}
\newcommand{\uq}{\mathrm{q}} \newcommand{\uQ}{\mathrm{Q}}
\newcommand{\ur}{\mathrm{r}} \newcommand{\uR}{\mathrm{R}}
\newcommand{\us}{\mathrm{s}} \newcommand{\uS}{\mathrm{S}}
\newcommand{\ut}{\mathrm{t}} \newcommand{\uT}{\mathrm{T}}
\newcommand{\uu}{\mathrm{u}} \newcommand{\uU}{\mathrm{U}}
\newcommand{\uv}{\mathrm{v}} \newcommand{\uV}{\mathrm{V}}
\newcommand{\ux}{\mathrm{x}} \newcommand{\uX}{\mathrm{X}}
\newcommand{\uy}{\mathrm{y}} \newcommand{\uY}{\mathrm{Y}}
\newcommand{\uz}{\mathrm{z}} \newcommand{\uZ}{\mathrm{Z}}

\newcommand{\re}{\mathrm{r}}
\newcommand{\im}{\mathrm{i}}
\renewcommand{\Re}{\operatorname{Re}}
\renewcommand{\Im}{\operatorname{Im}}

\newcommand{\Nbbm}{\mathbbm{N}}
\newcommand{\Qbbm}{\mathbbm{Q}}
\newcommand{\Rbbm}{\mathbbm{R}}
\newcommand{\Cbbm}{\mathbbm{C}}
\newcommand{\myownbbm}[1]{\text{\usefont{U}{bbm}{m}{n}#1}}
\newcommand{\identitymatrix}{\myownbbm{1}}
\newcommand{\zeromatrix}{0}
\newcommand{\unitybbm}{\identitymatrix}

\ifdefined\abs
  \renewcommand{\abs}[1]{\left|#1\right|}
\else
  \newcommand{\abs}[1]{\left|#1\right|}
\fi

\newcommand{\p}{\partial}

\newcommand{\Lieop}[1]{L_{#1}}
\newcommand{\Lieoptilde}[1]{\tilde{L}_{#1}}
\newcommand{\trafoop}[2]{\mathcal{D}_{#1}^{#2}}
\newcommand{\trafooptilde}[1]{\tilde{\mathcal{D}}_{#1}}

\newcommand{\Ham}{\hat{H}}
\newcommand{\HamDensity}{\mathcal{H}}
\newcommand{\Jacobian}{\mathcal{J}}
\newcommand{\EACT}{E^{\ddagger}}

\newcommand{\eps}{\varepsilon}
\newcommand{\generator}{g}
\newcommand{\defined}{\definition} \newcommand{\definition}{\equiv}

\newcommand{\strich}[1]{\left.#1\right|}

\newcommand{\Ng}{{N_{\!\text{g}}}}
\newcommand{\dmn}{\delta^{mn}}
\newcommand{\dop}{\delta^{op}}
\newcommand{\armn}{\alpha_r^{mn}}
\newcommand{\arop}{\alpha_r^{op}}
\newcommand{\axmn}{\alpha_x^{mn}}
\newcommand{\aymn}{\alpha_y^{mn}}
\newcommand{\azmn}{\alpha_z^{mn}}
\newcommand{\azop}{\alpha_z^{op}}
\newcommand{\dmnop}{\delta^{mnop}}
\newcommand{\armnop}{\alpha_r^{mnop}}
\newcommand{\axmnop}{\alpha_x^{mnop}}
\newcommand{\aymnop}{\alpha_y^{mnop}}
\newcommand{\azmnop}{\alpha_z^{mnop}}
\newcommand{\kxmnop}{\kappa_x^{mnop}}
\newcommand{\kymnop}{\kappa_y^{mnop}}
\newcommand{\amn}{\alpha^{mn}}
\newcommand{\aop}{\alpha^{op}}
\newcommand{\amnop}{\alpha^{mnop}}
\newcommand{\EllipticRD}{\mathcal{R}_D}
\newcommand{\aad}{a}
\newcommand{\Emf}{E_\text{mf}}
\newcommand{\Tc}{T_\text{c}}
\newcommand{\kB}{k_\text{B}}
\newcommand{\transpose}{\mathsf{T}}
\newcommand{\symplecticJ}{\mathcal{J}}
\newcommand{\phasespacevec}{u}
\newcommand{\dof}{{{d}}}
\newcommand{\drrr}{\ud^3r}
\newcommand{\manifold}{\mathcal{P}}
\newcommand{\manifoldQ}{\mathcal{Q}}
\newcommand{\manifoldN}{\mathcal{N}}
\newcommand{\configspace}{\manifoldQ}

\newcommand{\zr}{z_{\re}}
\newcommand{\zi}{z_{\im}}

\newcommand{\Htilde}{\tilde{H}}
\newcommand{\Hhat}{\hat{H}}
\newcommand{\Hdensity}{\mathcal{H}}
\newcommand{\nmax}{{n_\mathrm{max}}}
\newcommand{\nNFO}{{n_\mathrm{NFO}}}
\newcommand{\ad}[1]{\operatorname{ad}_{#1}}
\newcommand{\Mad}[1]{\operatorname{Mad}_{#1}}
\newcommand{\smatrix}{S}

\newcommand{\Lie}[2]{\left[#1,#2\right]}
\newcommand{\Dop}[2]{\mathcal{D}_{\!#1}^{#2}}
\newcommand{\Lop}[2]{\mathcal{L}_{\!#1}^{#2}}

\newcommand{\Mset}{\mathcal{M}}

\newcommand{\fluxfunction}{\mathcal{F}}
\newcommand{\phasespacevolflux}{\mathcal{F}^*}
\newcommand{\flux}{\mathcal{F}^*}

\newcommand{\Ball}{\mathcal{B}}
\newcommand{\Sphere}{\mathcal{S}}

\newcommand{\partitionfunction}{\mathcal{Z}}
\newcommand{\UNF}{U}

\newcommand{\mathcomment}[1]{\text{\textcolor{orange}{\texttt{[#1]}}}}

\newcommand{\order}[1]{\mathcal{O}\!\left({#1}\right)}
\newcommand{\orderof}{\operatorname{\sim}}

\newcommand{\poissonbracket}[1]{\left\lbrace #1 \right\rbrace}

\newcommand{\cpop}[2]{\cfrac{\partial #1}{\partial #2}}

\newcommand{\unitvec}{\hat\ee}

\newcommand{\Vext}{V_\text{ext}}
\newcommand{\Vint}{V_\text{int}}
\newcommand{\fluc}{\hat\delta}

\newcommand{\acrit}{a_\text{crit}}
\newcommand{\apf}{a_\text{pb}}
\newcommand{\abif}{a_\text{bif}}

\newcommand{\qx}{q_x}
\newcommand{\qy}{q_y}
\newcommand{\qz}{q_z}

\renewcommand{\propto}{\sim}

\newcommand{\Hlin}{\hat{H}_\text{l}}
\newcommand{\Hnonlin}{\hat{H}_\text{nl}}

\newcommand{\kmaxL}{k_{\text{max},\Lop{}{}}}
\newcommand{\kmaxD}{k_{\text{max},\Dop{}{}}}
\newcommand{\jmax}{j_\text{max}}

\newcommand{\hilfsnorm}{\mathcal{S}}
\newcommand{\hilfsenergie}{\mathcal{U}}
\newcommand{\Etilde}{\mathcal{E}}

\newcommand{\nonneg}{\text{nonneg.}}

\newcommand{\summea}{\sigma_{\mm (2i-1)}}
\newcommand{\summec}{\sigma_{\mm (2i)}}
\newcommand{\summeaj}{\sigma_{\mm (2j-1)}}
\newcommand{\summecj}{\sigma_{\mm (2j)}}

\newcommand{\Nres}{N_\text{res}}
\newcommand{\Neq}{N_\text{eq}}
\newcommand{\nbin}{n_\text{bin}}
\newcommand{\BEC}{BEC}

\newcommand{\eq}{Eq.}
\newcommand{\Eq}{\eq}
\newcommand{\EQ}{\eq}

\newcommand{\eqs}{Eqs.}
\newcommand{\Eqs}{\eqs}
\newcommand{\EQS}{\eqs}

\newcommand{\seq}{Sec.}
\newcommand{\Sec}{\seq}
\newcommand{\SEC}{\seq}

\newcommand{\seqs}{Secs.}
\newcommand{\Secs}{\seqs}
\newcommand{\SECS}{\seqs}

\newcommand{\fig}{Fig.}
\newcommand{\Fig}{\fig}
\newcommand{\FIG}{\fig}

\newcommand{\figs}{Figs.}
\newcommand{\Figs}{\figs}
\newcommand{\FIGS}{\figs}

\newcommand{\tab}{Tab.}
\newcommand{\Tab}{\tab}
\newcommand{\TAB}{\tab}

\newcommand{\Ref}{Ref.}
\newcommand{\REF}{\Ref}

\renewcommand{\refs}{Refs.}
\newcommand{\Refs}{\refs}
\newcommand{\REFS}{\refs}

\newcommand{\chap}{Chap.}
\newcommand{\Chap}{\chap}
\newcommand{\CHAP}{\chap}

\newcommand{\chaps}{Chaps.}
\newcommand{\Chaps}{\chaps}
\newcommand{\CHAPS}{\chaps}

\newcommand{\app}{Appx.}
\newcommand{\App}{\app}
\newcommand{\apps}{Appxs.}
\newcommand{\Apps}{\apps}
\newcommand{\APPS}{\apps}

\newcommand{\eg}{e.\;\!g.}
\newcommand{\EG}{\eg.}
\newcommand{\Eg}{E.\;\!g.}
\newcommand{\ie}{i.\;\!e.}
\newcommand{\IE}{\ie}
\newcommand{\cf}{cf.}
\newcommand{\CF}{\cf.}

\newcommand{\new}[1]{\textcolor{blue}{#1}}

\title[Construction of Darboux coordinates and Poincar\'e-Birkhoff normal forms 
       \ldots]
      {Construction of Darboux coordinates and Poincar\'e-Birkhoff normal forms 
       in noncanonical Hamiltonian systems}
\author{Andrej Junginger, Jörg Main, and Günter Wunner}
\address{1. Institut f\"{u}r Theoretische Physik, 
         Universit\"{a}t Stuttgart, 
         70550 Stuttgart, Germany}
\date{\today}

\begin{abstract}
We demonstrate a general method to construct Darboux coordinates via normal form 
expansions in noncanonical Hamiltonian system obtained from \eg\ a variational 
approach to quantum systems.
The procedure serves as a tool to naturally extract canonical coordinates out of 
the variational parameters and at the same time to transform the energy 
functional into its Poincar\'{e}-Birkhoff normal form.
The method is general in the sense that it is applicable for arbitrary degrees 
of freedom, in arbitrary orders of the local expansion, and it is independent of 
the precise form of the Hamilton operator.
The method presented allows for the general and systematic investigation of 
quantum systems in the vicinity of fixed points, which \eg\ correspond to 
ground, excited or transition states. Moreover, it directly allows to calculate 
classical and quantum reaction rates by applying transition state theory.
\end{abstract}

\section{Introduction}

It is at the core of physical sciences to describe and investigate the dynamics 
of systems. Depending on their nature, these can either be described by 
the Schrödinger equation in case of quantum mechanical systems, or \eg\ in terms 
of Hamiltonian mechanics in case of a classical system. 
In both cases, a canonical structure of the dynamical equations 
\cite{Rebhan1999,Rebhan2005} is inherent 
which is expressed in the existence of conjugate pairs of field operators 
$\hat\psi, \hat\psi^\dagger$ with infinite degrees of freedom or conjugate 
coordinates $\qq, \pp$ with a finite number of degrees of freedom. Both 
approaches serve as powerful frameworks to investigate a huge amount of 
different physical problems.
In addition to the global dynamics of a physical system which can be determined 
by solving the corresponding equations of motion, its fixed points play a 
crucial role in many investigations: 
For example, fixed points which correspond to a (local) minimum of the 
Hamiltonian form (metastable) ground states of the system. 
Moreover, fixed points which are related to saddle points of the Hamiltonian are 
unstable, excited states. A special class of such unstable fixed points are 
rank-1 saddle points which possess exactly one unstable direction. These points 
are of special interest in dynamical systems, because they form bottlenecks in 
the underlying phase space which separate different regions therein. Considering 
a dynamical system, the transition from one to the other subregion of phase 
space is then mediated by the saddle point. Therefore, the latter determines the 
reaction dynamics between the different subregions which is the basic statement 
of transition state theory 
\cite{Langer1967, Langer1969, pitzer, truh79, pechukas1981,%
Laidler1983, truh85, Haenggi1990, truhlar91, truh96, truh2000,%
Komatsuzaki2001, Waalkens2008, hern08d, Komatsuzaki2010, hern10a,%
Henkelman2016}.

Beyond the fixed points of the system's dynamical equations themselves, their 
local properties are of high interest in many applications. For example, the 
local properties of a minimum of the Hamiltonian determine the physics of the 
system for small excitations from the ground state. Moreover, the local 
properties in the vicinity of a rank-1 saddle point or transition state 
determine the reaction dynamics and rates of the system.

\EDITS{
For a detailed analysis of the local fixed point properties of a canonical 
Hamiltonian system \STRIKE{as well as for the construction of a dividing 
surface that 
separates reactants and products in it,} a standard and powerful tool is its 
normal form expansion \cite{Waalkens2008, Murdock2010, Kawai2009a}. 
Especially in the field of reaction dynamics, the normal form Hamiltonian in 
the vicinity of rank-1 saddle points is important, because it provides a way of 
defining a normally hyperbolic invariant manifold
\cite{bristol2,pollak78,pech79a,hern93a,%
hern93b,hern94,Jaffe05,Komatsuzaki99,Uzer02,Waalkens04b,Li06prl,Waalkens2008,%
Kawai2009a,hern10a,Teramoto11,Waalkens13}
with which a nonrecrossing dividing surface between reactants and products in 
multi-degree-of-freedom systems can be constructed.
}

\EDITS{
In \REF~\cite{Waalkens2008}, Waalkens \etal describe this procedure in detail, 
of which our work will be a natural extension to \emph{noncanonical} 
coordinates.
We therefore give a brief overview of the method in the following:
If the Hamiltonian $H$ is given in terms of a set of canonical coordinates 
$\qq,\pp$ then its normal form can be constructed via the following expansion,
\begin{equation}
   \tilde H (\qq,\pp)
   = \sum_{j=0}^\infty
      \frac{1}{j!} \,
      \ad{W}^j H (\qq,\pp) \,.
\label{eq:NF-H-CNF}
\end{equation}
Here, $W$ is an appropriate generating function and $\ad{W} H = \lbrace W, 
H\rbrace$ is the adjoint operator that equals the definition of the Poisson 
bracket.
Usually, the normal form Hamiltonian is required up to a certain polynomial 
order within a local expansion at a fixed point.
Consequently, it is appropriate to regard, in general, expansions of all 
quantitues occurring in \EQ~\eqref{eq:NF-H-CNF}, \ie~the original Hamiltonian, 
the transformed one, and the generating function.
This procedure has the advantage that the transformation in  
\EQ~\eqref{eq:NF-H-CNF} can be applied order by order.
Waalkens \etal~\cite{Waalkens2008} describe in detail how these single steps 
are performed and how exactly the generating function $W$ needs to be 
constructed through a homological equation in order to obtain the 
Poincar\'e-Birkhoff normal form of the Hamilton function (we refer the reader to 
this reference for more details).
The final result is then, by construction, a Hamiltonian $H(\JJ)$ which depends 
on the actions coordinates $\JJ$ all being constants of motion up to the 
respective order of the expansions.
}

\EDITS{
With special regard to reactive systems, this normal form is of particular 
advantage, because -- if $J_1$ corresponds to the reaction channel of the 
system, \ie\ the unstable direction of a rank-1 saddle point -- then a local,
recrossing-free dividing surface is defined by $J_1=0$,  
}
The directional flux through the dividing surface at fixed energy $E$ is then
given by 
\begin{equation}
  f(E) = (2\pi)^{\dof-1} \, \mathcal{V}(E) \,,
  \label{eq:directional-flux}
\end{equation}
where $\mathcal{V}(E)$ is the volume of actions $(J_2,\ldots,J_\dof)$ enclosed 
by the contour $H(0,J_2,\ldots,J_\dof)=E$ and 
the thermal reaction rate $\Gamma$ at temperature $\beta=1/\kB T$ is 
obtained from the Boltzmann average of \EQ\ \eqref{eq:directional-flux} which 
yields (\cf\ \REF\ \cite{toller})
\begin{equation}
  \Gamma =
  \frac{1}{2\pi \beta}
  \frac{
    \int \ud J_2 \ldots \ud J_\dof \, 
    \exp \bigl(-\beta H(0, J_2, \ldots, J_\dof)  \bigr) 
  }{
    \int \ud J'_1 \ldots \ud J'_\dof \, 
    \exp \bigl(-\beta H'(J'_1, \ldots, J'_\dof) \bigr) 
  }\,.
\label{eq:thermal-rate}
\end{equation}
Here, $H$ is the normal form at the transition state and $H'$ that at the 
metastable ground state.
In this context of reaction dynamics the importance of the knowledge of a 
normal form Hamiltonian and, in order to be able to actually calculate reaction 
rates, an explicit construction scheme of the local action variables become 
obvious. 
\EDITS{
We note that the work of Waalkens \etal~\cite{Waalkens2008} goes even beyond 
this by also introducing how quantum reaction rates can be calculated within a 
formally equivalent procedure that merely requires a redefinition of the 
adjoint operator.
}
\STRIKE{
In case of a classical Hamiltonian system which is described in terms of 
canonical coordinates $\qq,\pp$, such a scheme is in detail described in 
\REF~[15].
}

\EDITS{
It is the purpose of this paper to extend the scheme of Waalkens 
\etal~\cite{Waalkens2008} to the more general field of \emph{noncanonical} 
Hamiltonian systems, as \eg\ quantum mechanical wave packets whose dynamics is 
governed by the Schrödinger equation (see below).
Therefore, we will describe in the following a quantum system within a 
variational approach, determine the respective dynamical equations by applying 
a time-dependent variational principle \cite{Frenkel1934,McLachlan1964}, and 
show that it defines a general, \emph{noncanonical} Hamiltonian system (see 
below for a precise definition). 
}
In such quantum systems, fixed points of the dynamical equations and their 
local properties have the same meaning for the quantum reaction dynamics as they 
have in classical systems. 
For a detailed analysis of the local properties, it is, therefore, desirable to 
obtain an analogue of the classical normal form also for the quantum system.
However, the usual treatment \eqref{eq:NF-H-CNF} cannot be applied, because 
neither a classical Hamilton function $H(\qq,\pp)$ in canonical coordinates nor 
such coordinates themselves are known.

Here, we present a method by which both the transformation of the 
variational approach as a noncanonical Hamiltonian system into its 
Poincar\'e-Birkhoff normal form and simultaneously the construction of canonical 
coordinates is obtained. 
The result of the transformations is, by construction, a set of canonical normal 
form coordinates. In the latter, the energy functional of the system will serve 
as a classical Hamilton function which has the advantageous property that it is 
directly formulated in action variables. If truncated at a certain order, the 
constructed Hamiltonian will serve as an approximation to the true quantum 
system which directly allows for the application of transition state theory and 
the evaluation of quantum reaction rates via \EQS\ \eqref{eq:directional-flux} 
and \eqref{eq:thermal-rate}.
\EDITS{
In technical terms, the crucial difference between our procedure in 
noncanonical coordinates and the usual treatment in canonical ones is that we 
treat the dynamical equations as well as the energy functional separately.
From the mathematical point of view, this brings with it that the generating 
function of the transformation and the corresponding operators require a 
different definition than in \EQ~\eqref{eq:NF-H-CNF}.
}

Our paper is organized as follows:
In \SEC\ \ref{sec:TDVP} we introduce a variational approach to quantum systems 
which defines a noncanonical Hamiltonian system for the variational parameters. 
Furthermore, we discuss its formal relation to classical canonical mechanics 
and some important fixed point properties of the linearised dynamical equations.
In \SEC\ \ref{sec:construction}, the method to construct local canonical 
coordinates in the vicinity of the fixed point is introduced. Therefore, a 
symplectic basis formed by appropriately normalized eigenvectors of the 
linearised dynamical equations is used and higher-order terms of the expansions 
are treated via normal form transformations. As a key feature -- and in contrast 
to the usual transformation  \eqref{eq:NF-H-CNF} of canonical Hamiltonians -- 
this procedure treats the dynamical equations and the energy functional 
separately. Moreover, the normal form expansions are carried out in two steps: 
First, its polynomial structure is generated using the nonresonant terms of the 
corresponding generating function (see below for the latters' definition). 
Second, the remaining resonant coefficients of the generating function which are 
free parameters are chosen in such a way that the dynamical equations as well as 
the energy functional in normal form coordinates fulfil canonical equations, 
\ie\ the normal form coordinates are canonical ones by construction.
\EDITS{
We have written the paper such that the essential steps that go 
beyond the work in \REFS~\cite{Waalkens2008, Murdock2010} are presented in the 
\ref{last-theorem} theorems presented in \SEC~\ref{sec:construction}.
In the appendix, we provide in addition both a numerical example of the 
presented procedure and an exemplary \textsc{Mathematica} script code, in which 
the reader is welcome to execute the respective steps while reading the paper.
}

We note that it is not within the scope of this paper to deal with questions of 
existence and convergence of the objects made use of, but to present a scheme 
analogously to and beyond \REF\ \cite{Waalkens2008} by which canonical 
coordinates and the normal form can be constructed at the same time. 
The method developed in this paper presents the basis of \eg\ the calculation of 
thermal decay rates of a metastable 1-dimensional potential as well as 
Bose-Einstein condensates with different kinds of interactions. Results using 
the leading order and including higher orders of the normal form expansion are 
presented in Refs.\ \cite{Junginger2012d, Junginger2013b} and 
\cite{Junginger2012a, Junginger2012b, Junginger2015a}, respectively.

\section{Variational approach to quantum systems as a noncanonical Hamiltonian 
system}
\label{sec:TDVP}

In this paper, we focus on quantum systems which are described by the 
Schrödinger equation
\begin{equation}
  \ui \hbar \pop{}{t} \, \psi(\rr,t) = \hat H \psi(\rr,t) \,.
  \label{eq:QM-Schroedinger-equation-timedep}
\end{equation}
Here, $\psi(\rr,t)$ is the time-dependent wave function of the system and $\hat 
H$ is the Hamilton operator. As it is well known, there is a natural canonical 
structure inherent to this description. This becomes especially obvious, if one 
derives the Schrödinger equation in the framework of field theory from the 
Hamiltonian density
\begin{equation}
  \mathcal{H} =
  \int \drrr ~
  \psi^\dagger (\rr,t) \, \hat H \, \psi(\rr,t) 
\end{equation}
using the functional derivatives
\begin{equation}
  \ui \hbar \pop{}{t} \psi(\rr,t) =
  \pop{\mathcal{H}}{\psi^\dagger (\rr,t)}
  \,, \qquad
  \ui \hbar \pop{}{t} \psi^\dagger(\rr,t) =
  - \pop{\mathcal{H}}{\psi (\rr,t)} \,.
\end{equation}
This description of a quantum system is very general, however, it is often not 
feasible in actual applications due to the field operator's infinite number of 
degrees of freedom.

One possible approach to reduce the system's number of degrees of freedom is 
its treatment within a variational approach. Therein, the Schrödinger equation 
\eqref{eq:QM-Schroedinger-equation-timedep} is solved approximately by 
replacing the original wave function $\psi(\rr,t)$ by a trial wave function
\begin{equation}
   \psi(\rr,t) = \psi(\rr, \zz(t)) \,. 
   \label{eq:TDVP-trial-wave-function}
\end{equation}
Here, $\zz(t) = [z_1(t), z_2(t), \ldots, z_\dof(t)]^\transpose \in \Cbbm^\dof$ 
is a set of complex and time-dependent variational parameters, and the time 
evolution of the wave function is completely determined by that of the 
variational parameters.
In the framework of the variational approach, expectation values of the 
system's 
observables, in general, become functions depending on the variational 
parameters $\zz(t)$. 
In particular, the energy functional of the system is given by the expectation 
value of the Hamilton operator
\begin{equation}
  E(\zz(t)) = 
  \bra[big]{\psi(\rr,\zz(t))} 
    \hat H
  \ket[big]{\psi(\rr,\zz(t))} \,. 
  \label{eq:TDVP-energy-functional}
\end{equation}
In order to describe the dynamics of the system in the Hilbert subspace which is 
spanned by the variational ansatz, we apply the Dirac-Frenkel-McLachlan 
variational principle \cite{Frenkel1934, McLachlan1964}. This claims to minimize 
the norm of the difference between the left- and the right-hand side of the 
Schrödinger equation \eqref{eq:QM-Schroedinger-equation-timedep},
\begin{equation}
   I 
   \definition 
   \norm{ \ui \phi - \hat{H} \psi }^2
   =
   \braket[big]{ - \ui \phi - \hat{H} \psi }
               {   \ui \phi - \hat{H} \psi }
   \stackrel{!}{=} \text{min.} 
   \label{eq:McLachlan-variational-principle}
\end{equation}
Here, $\hbar = 1$ has been set, the arguments of the wave function $\psi$ have
been omitted for brevity, and also the time dependence of the variational
parameters $\zz$ will be dropped in the following.
The quantity $I$ is minimized with respect to $\phi$ and $\phi \defined 
\dot\psi$ is set afterwards which means that the Schrödinger equation is solved 
within the Hilbert subspace of the variational ansatz with the least possible 
error.
Since the approximate solution of the Schrödinger equation is intended to
minimize the quantity $I$, the latter's variations must vanish,
\begin{equation}
   \delta I = \braket[big]{ - \ui\, \delta \phi }{ \ui \phi - \hat{H} \psi} +
              \braket[big]{ - \ui   \phi - \hat{H} \psi}{ \ui \, \delta \phi }
            \stackrel{!}{=} 0 \,. 
   \label{eq:TDVP-I-variation}
\end{equation}
Because of \EQ\ \eqref{eq:TDVP-trial-wave-function},  the time derivative of the
trial wave function, $\phi = \dot \psi$, and its variation $\delta \phi$ yield
\begin{align}
   \phi = \sum_{m=1}^\dof \pop{\psi}{z_m} \, \dot{z}_m \,, \qquad
   \delta\phi = \sum_{n=1}^\dof \pop{\psi}{z_n} \, \delta \dot{z}_n \,, 
\end{align}
so that one obtains
\begin{equation}
   \delta I = \sum_{m,n=1}^\dof
              \braket[bigg]
                     { \pop{\psi}{z_m} }
                     { - \pop{\psi}{z_n} \dot{z}_n - \ui \hat{H} \psi}
                     \delta \dot{z}^*_m +
              \braket[bigg]
                     { - \pop{\psi}{z_n} \dot{z}_n + \ui \hat{H} \psi }
                     { \pop{\psi}{z_m} }
                     \delta \dot{z}_m
            \stackrel{!}{=} 0 \,. 
    \label{eq:variation-I}
\end{equation}
We now proceed from the complex variational parameters $\zz$ to their real and 
imaginary parts 
\begin{equation}
  \xx \defined (\zz^\re, \zz^\im)^\transpose \in \Rbbm^{2\dof}\,. 
  \label{eq:def-xx}
\end{equation}
In this case, the variations with respect to the variational parameters in \EQ\ 
\eqref{eq:variation-I} are not independent, and both terms together result in 
the dynamical equations 
\begin{equation}
  \sum_{n=1}^{2\dof}~
   \text{Im}\braket[bigg]
          { \pop{\psi}{x_m} }
          { \pop{\psi}{x_n} } \dot{x}_n
   = - \text{Re}
   \braket[bigg]
          { \pop{\psi}{x_m} }
          { \Ham \psi } 
   \label{eq:TDVP-equations-of-motion-real}
\end{equation}
for the time evolution of each real variational parameter $m=1,\ldots, 2\dof$.
Using the property
\begin{align}
  \pop{}{\xx}
  E(\xx)
  &=
  2 \Re \bra[bigg]{\pop{\psi(\xx)}{\xx}} \hat H \ket[bigg]{\psi(\xx)} \,,
\label{eq:TDVP-deriv-of-E}
\end{align}
which directly follows from \EQ\ \eqref{eq:TDVP-energy-functional} with the 
replacement \eqref{eq:def-xx} and the definitions
\begin{subequations}
\begin{align}
    K_{mn} &\definition 
   2 \Im \braket[bigg]
          { \pop{\psi}{x_m} }
          { \pop{\psi}{x_n} } \,, 
   \label{eq:TDVP-K-real} \\
    h_m &\definition
   2 \Re
   \braket[bigg]
          { \pop{\psi}{x_m} }
          { \Ham \psi } \,, 
   \label{eq:TDVP-h-real}
\end{align}%
\label{eq:TDVP-K-h-real}%
\end{subequations}
the dynamical equations \eqref{eq:TDVP-equations-of-motion-real} immediately 
take the form 
\begin{equation}
  K(\xx) \, \dot{\xx}
  = - \pop{E(\xx)}{\xx}
  \defined - \vec{h}(\xx) \,,
\label{eq:TDVP-equations-of-motion-real-short-Ham-form}
\end{equation}
which will be the basis of all considerations in this paper.
We note that the matrix $K$ with the entries \eqref{eq:TDVP-K-real} is 
skew-symmetric by definition, because the imaginary part changes its sign under 
complex conjugation of the bracket. 
Therefore, $K$ induces a symplectic geometry onto the space of variational 
parameters that can be expressed by the corresponding 2-form
\begin{align}
   \omega^2 =
   \sum_{\substack{m,n=1 \\ m<n}}^{2\dof}
   K_{mn} (\xx) \,
   \ud x_m \wedge \ud x_n  \,. 
   \label{eq:TDVP-2-form-K-tilde}
\end{align}
This 2-form is nondegenerate if $K$ is invertible which we will assume 
throughout this paper. Moreover, it is closed, \ie\ its exterior derivative 
vanishes,
\begin{align}
   \ud \omega^2
   =
   \sum_{\substack{m,n,k=1 \\ m<n}}^{2\dof}
   \pop{K_{mn}(\xx)}{x_k} \;
   \ud x_k \wedge \ud x_m \wedge \ud x_n
   = 0 \,,
  \label{eq:TDVP-2-form-ext-deriv}
\end{align}
because  the single terms 
\begin{equation}
  \p_{x_k} K_{mn} = 
    \Im \braket[bigg]{\frac{\p^2 \psi}{\p x_k \p x_m}}{\frac{\psi}{\p x_n}} -
    \Im \braket[bigg]{\frac{\p^2 \psi}{\p x_k \p x_n}}{\frac{\psi}{\p x_m}} \,,
\end{equation}
cancel out when it is summed over $k,m,n$.
Under these conditions Darboux's theorem \cite{Darboux1882, Arnold1989} 
guarantees the existence of local canonical coordinates. 
In \SEC\ \ref{sec:construction} we present a scheme by which such coordinates 
can actually be constructed via normal form expansions.

\subsection{Relation to classical canonical Hamiltonian mechanics}
\label{sec:noncan-Ham-sys}

The dynamical equations \eqref{eq:TDVP-equations-of-motion-real-short-Ham-form} 
are formally equivalent to Hamilton's equations in classical mechanics. To see 
this, let us consider a Hamiltonian system with $\dof$ degrees of freedom, 
whose standard canonical coordinates $\vec\phasespacevec = (q_1, p_1, 
\ldots, q_\dof, p_\dof)^\transpose$, fulfil the Poisson brackets 
\begin{equation}
  \left\lbrace q_i, p_j \right\rbrace=\delta_{ij} \,, \quad
  \left\lbrace q_i, q_j \right\rbrace=0 \,, \quad
  \left\lbrace p_i, p_j \right\rbrace=0 \,.
  \label{eq:Poisson-brackets}
\end{equation}
The physics of this system is described by the Hamiltonian 
$H=H(\vec\phasespacevec)$ and the dynamics of the system is then determined by 
Hamilton's equations
\begin{equation}
   \symplecticJ \dot{\vec{\phasespacevec}} =
   - \pop{H(\vu)}{\vec \phasespacevec}  \,, 
   \label{eq:Ham-Eqs-gamma-Jlinks}
\end{equation}
where, $\symplecticJ$ is the standard symplectic matrix
\begin{equation}
   \symplecticJ \defined
   \begin{pmatrix}
      \symplecticJ_1 & & 0 \\
      & \ddots & \\
      0 & & \symplecticJ_1 
   \end{pmatrix} \,, 
\qquad \text{with} \qquad
   \symplecticJ_1 \defined
   \begin{pmatrix}
      0 & 1\\
      - 1& 0
   \end{pmatrix} \,. 
   \label{eq:Ham-symplecticJ-definition}
\end{equation}
The latter relates the Hamiltonian vector field $\dot\vu$ to the derivative 
$\partial H / \partial \vu$ of the Hamiltonian, and it induces a symplectic 
geometry on phase space.

In the dynamical equations  
\eqref{eq:TDVP-equations-of-motion-real-short-Ham-form}, the time-derivative 
$\dot\xx$ is also related to the gradient of the energy functional $E$ via a 
skew-symmetric matrix, here $K$. The crucial difference to \EQ\ 
\eqref{eq:Ham-Eqs-gamma-Jlinks} is that, because of its definition according to 
\EQ\ \eqref{eq:TDVP-K-real}, $K$ has a more complicated structure than 
$\symplecticJ$, in particular it generally depends on the variational 
parameters. 
As a consequence, the Poisson brackets \eqref{eq:Poisson-brackets} are not 
fulfilled for the variational parameters in the system 
\eqref{eq:TDVP-equations-of-motion-real-short-Ham-form} and, for this reason, 
we will refer to the latter as a \emph{noncanonical} Hamiltonian system in this 
paper.
The fact that the matrix $K$ is not equal to $\symplecticJ$ also leads to the 
fact that one can no longer identify certain pairs of ``conjugate coordinates'' 
$x_i, x_j$ in variational space, because \emph{all} the time-derivatives $\dot 
x_i, \dot x_j$ ($i,j=1,\ldots,\dof$) are, in general, coupled in a nontrivial 
way.

Both dynamical equations \eqref{eq:TDVP-equations-of-motion-real-short-Ham-form} 
and \eqref{eq:Ham-Eqs-gamma-Jlinks} have in common that they describe a 
\emph{classical} dynamics. However, we emphasize that the classical dynamics 
\eqref{eq:TDVP-equations-of-motion-real-short-Ham-form} takes place in 
variational space and that this is an effective description of the fully 
quantized physical system that is described by the Schrödinger equation 
\eqref{eq:QM-Schroedinger-equation-timedep}. Consequently, there is no need to 
take into account a further quantization of the variational parameters 
(including possible problems in context with their nature as noncanonical 
coordinates).

\subsection{Local dynamical equations and their eigenvalue structure}

As already mentioned in the introduction, we focus on the local dynamics of \EQ\ 
\eqref{eq:TDVP-equations-of-motion-real-short-Ham-form}  in the vicinity of a 
fixed point. Thus, in the following, we consider local Taylor expansions of $K$ 
and $\hh$ at the fixed point up to any desired order $\nmax$,
\begin{equation}
  \left( \sum_{n=0}^{\nmax-1} K_n(\xx) \right) \dot \xx
  = - \sum_{n=1}^{\nmax}  \hh_n (\xx) \,,
   \label{eq:NF-DGL-expansion-separately}
\end{equation}
where the matrix $ K$ and the vector $ \hh$ are expanded independently according 
to
\begin{subequations}
\begin{align}
   K (\xx)
   &\approx \sum_{n=0}^{\nmax-1} K_n (\xx)
   \,,  \\
   \hh (\xx)
   &\approx \sum_{n=1}^\nmax \hh_n (\xx)
   \,. 
\end{align}%
\label{eq:NF-multivariate-expansion-K-h}%
\end{subequations}
Analogously, the energy functional is expanded as 
\begin{equation}
  E(\xx) \approx \sum_{n=0}^{\nmax+1} E_n (\xx) \,.
  \label{eq:NF-E-expansion}
\end{equation}
The terms $K_n$, $\hh_n$, and $E_n$ summarize all terms of the respective 
expansion which are homogeneous of degree $n$, and $\hh_0=0$ vanishes because 
the expansion is performed at a fixed point. 
Alternatively, the expansion of the equations of motion 
\eqref{eq:NF-DGL-expansion-separately} can be rewritten equivalently in the form
\begin{equation}
  \dot \xx 
  = -  K^{-1}(\xx) \;  \hh(\xx)
  \approx \sum_{n=1}^\nmax \ff_n (\xx) \,,
   \label{eq:NF-DGL-starting}
\end{equation}
where both $K$ and $ \hh$ are combined on the same side of the equation and 
where $\ff_n$ collects the terms of order $n$. 

For the following considerations, the local eigenvalue structure of the 
dynamical equations at a fixed point $\dot\xx=0$ are of fundamental 
importance. These are determined by the linearised dynamical equations
\begin{equation}
   K_0 \, \dot \xx = -  \hh_1 (\xx) = F \, \xx \,,
  \label{eq:TDVP-DGL-linearized-separately}
\end{equation}
where it is assumed that the fixed point is located at the origin $\xx=0$ for 
simplicity (this can always be achieved by a simple shift of the coordinates).
$ K_0$ is the zeroth-order expansion of the matrix $ K$ and $\hh_1 = - F 
\xx$ is the linearised vector $\hh$ at the fixed point.
Because $ K$ is skew-symmetric in general, this property of course also holds 
for its zeroth-order approximation. The matrix $F$ is symmetric, because it is 
the negative Hessian matrix of the energy functional according to \EQ\ 
\eqref{eq:TDVP-equations-of-motion-real-short-Ham-form}, $F_{mn} = -\partial^2 
E/\partial x_m \partial x_n$. Consequently, the equations
\begin{equation}
  K_0 = -K_0^\transpose\,, \qquad F = F^\transpose 
  \label{eq:TDVP-symmetry-K0-F}
\end{equation}
hold. In order to obtain the eigenvalue spectrum of the linearised equations of
motion, the first-order differential equation
\eqref{eq:TDVP-DGL-linearized-separately} is solved using the ansatz $\xx(t) = 
\vv \, \ue^{\lambda t} $, where $\lambda \in \Cbbm$ is a complex 
parameter, and $\vv \in \Cbbm^{2\dof}$ is a complex vector.
Inserting this ansatz into \EQ\ \eqref{eq:TDVP-DGL-linearized-separately}, one 
obtains the generalized eigenvalue equation
\begin{equation}
 F \, \vv = \lambda K_0 \, \vv \,.
\end{equation}
The eigenvalues $\lambda$ are the roots of the characteristic polynomial $\chi 
(\lambda)= \det (F - \lambda K_0)$, and with the properties 
\eqref{eq:TDVP-symmetry-K0-F}, it can easily be shown that the characteristic 
polynomial is an even function of $\lambda$, \ie\ $\chi(\lambda) = 
\chi(-\lambda)$.
Thus, if $\lambda$ is a root of the characteristic polynomial, then also 
$-\lambda$ is a root, so that all the eigenvalues occur pairwise with different 
sign.
Therefore, the eigenvalue spectrum of the linearised dynamical equations in the
vicinity of a fixed point always exhibits the structure
\begin{equation}
   \vec\lambda^\pm =
   ( +\lambda_1, 
     -\lambda_1, 
   \ldots, 
     +\lambda_\dof,
     -\lambda_\dof
     ) 
   \label{eq:TDVP-EV-structure-pm-structure}
\end{equation}
which will be of fundamental importance for the normal form expansions 
performed in the next \SEC\ \ref{sec:construction}.

\section{Construction of canonical normal form coordinates}
\label{sec:construction}

The knowledge of canonical coordinates is fundamental to many methods known 
from classical Hamiltonian mechanics and, beyond their existence, a central 
question is how these can be constructed.
As the key result of this paper, a general method to construct canonical normal 
form coordinates will be presented in this section. As will be shown, this 
method has the advantage that it simultaneously yields both a procedure to 
extract canonical coordinates and a transformation of the system into its 
Poincar\'{e}-Birkhoff normal form. 
The procedure consists of the following three steps (see \FIG\ 
\ref{fig:NF-principle}):

\begin{figure}[t]
\centering
\includegraphics[width=.6\columnwidth]{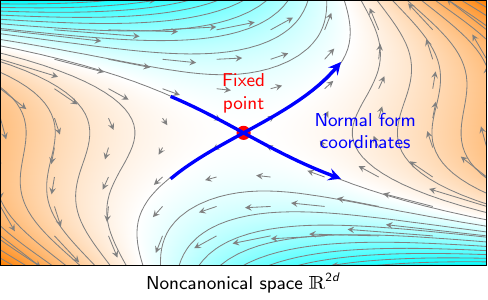}
\caption{%
Schematic illustration of the normal form coordinates.
The contour lines represent isosurfaces of the energy functional $E$ and the
arrows depict the vector field $\dot \xx$ obtained by the dynamical equations 
\eqref{eq:TDVP-equations-of-motion-real-short-Ham-form}.
In the vicinity of a fixed point (red circle), the normal form coordinates 
define a local coordinate system whose origin is the fixed point.
}
\label{fig:NF-principle}
\end{figure}

\begin{enumerate}[(i)]

\item
In the first step, the expansions \eqref{eq:NF-multivariate-expansion-K-h} and 
\eqref{eq:NF-E-expansion} are transformed via a linear change of coordinates to 
a symplectic basis which is defined by the eigenvectors of the linearised 
dynamical equations. The resulting diagonal coordinates are canonical ones in 
the first order of the expansions.

\item 
Successive Lie transforms are applied order by order to treat the higher-order 
corrections of the expansions. 
In the corresponding generating function two different types of terms will be 
distinguished, namely ``nonresonant'' and ``resonant'' coefficients (see below).
The generation of the normal form structure will be performed via the 
nonresonant terms, and all of them are determined uniquely by the requirement 
that certain monomials of the dynamical equations shall be removed.
Which of the terms remain after the Lie transforms is solely determined by a
resonance condition depending on the eigenvalues, and because of their general
structure \eqref{eq:TDVP-EV-structure-pm-structure}, the normal form 
will exhibit a fundamental polynomial structure.

\item 
The normal form expansions leave the freedom to choose the resonant terms of the
generating function.
In the last step, these free parameters are chosen in a way that the dynamical
equations and the energy functional fulfil canonical equations, \ie\ the normal
form coordinates are canonical ones by construction.

\end{enumerate}

\subsection{Symplectic basis}
\label{sec:symplectic-basis}
\label{sec:NF-step1}

\EDITS{
In order to ``simplify'' the system in its lowest order, it is sufficient to
focus on the linearised dynamical equations 
\eqref{eq:TDVP-DGL-linearized-separately}, and therein the following theorem 
holds:
}
\begin{thm}[symplectic basis]
\EDITS{
There exists a symplectic basis, within which the linearized dynamical equations 
\eqref{eq:TDVP-DGL-linearized-separately} can be transformed into a structure 
in which its left-hand side equals the standard symplectic matrix $\symplecticJ$ 
in \EQ~\eqref{eq:Ham-symplecticJ-definition} and its right-hand side possesses 
a block structure containing the eigenvalues of the linearized equations of 
motion.
Explicitely, there is a transformation matrix $T$ with the properties
\begin{align}
  T^\transpose K_0 T = \symplecticJ 
\qquad \text{and} \qquad
  T^\transpose F T =
  \begin{pmatrix}
    0 & \lambda_1 & & & \\
    \lambda_1 & 0 & & & \\
    & & \ddots & & \\
    & & & 0 & \lambda_\dof \\
    & & & \lambda_\dof & 0 
  \end{pmatrix} \,.
\end{align}
}
\end{thm}
\begin{proof}
\EDITS{
A natural basis of this linearised system is spanned by its eigenvectors 
$\vv_{2i-1}, \vv_{2i}$. These are solutions of the generalized eigenvalue 
problem
\begin{subequations}
\begin{align}
  F \, \vv_{2i-1} &= +\lambda_i \, K_0 \, \vv_{2i-1} \,, \\
  F \, \vv_{2i  } &= -\lambda_i \, K_0 \, \vv_{2i  } \,,
\end{align}
\end{subequations}
where $i=1,\ldots,\dof$.
To obtain the desired form, we normalize the eigenvectors by 
\begin{subequations}
\begin{gather}
    \bra{\vv_{2i-1}} K_0 \ket{\vv_{2i  }}  = 
  - \bra{\vv_{2i  }} K_0 \ket{\vv_{2i-1}}  = 1 \,,
\\
    \bra{\vv_{2i-1}} F \ket{\vv_{2i  }}  = 
    \bra{\vv_{2i  }} F \ket{\vv_{2i-1}}  = \lambda_i 
\end{gather}%
\label{eq:NF-eigenvectors-normalization}%
\end{subequations}
for all $i=1,\ldots,\dof$.
Combining the eigenvectors in the transformation matrix $T=(\vv_1, \ldots,
\vv_{2\dof})$, the choice \eqref{eq:NF-eigenvectors-normalization} by
construction guarantees the block structures
where $\symplecticJ$ is the standard symplectic matrix defined in \EQ\
\eqref{eq:Ham-symplecticJ-definition}. Consequently, the normalized eigenvectors
define a symplectic basis and the coordinates are canonical ones concerning the
linearised system.
}
\end{proof}

In order to regard the full, nonlinearised dynamical equations in this 
symplectic basis, the transformation $\xx\to\xx'=T^{-1}\xx$ needs to be applied 
also to the higher-order terms.
Emanating from \EQ\ \eqref{eq:NF-DGL-starting} and omitting the prime, this
linear change of coordinates transforms the dynamical equations into the form
\begin{equation}
  \dot \xx
  = T^{-1} \, \sum_{n=1}^\nmax \ff_n (T\xx)
  \defined \aa(\xx) = \sum_{n=1}^\nmax \aa_n (\xx) \,. 
  \label{eq:NF-DGL-starting-diagonalized-aa}
\end{equation}
In the last step, the single expansion coefficients have been redefined by the
coefficients $\aa_n$ which collect the terms homogeneous of degree $n$.
It is noted that, because the inverse matrix $T^{-1}$ is used here instead of
the transpose $T^\transpose$, the linear term of \EQ\
\eqref{eq:NF-DGL-starting-diagonalized-aa} is diagonal by construction,
\begin{equation}
  \aa_1 (\xx)
  = ( T^{-1} \, F \, T ) \, \xx
  = \begin{pmatrix}
      + \lambda_1 & & & & \\
      & - \lambda_1 & & & \\
      & & \ddots    & & \\
      & & & + \lambda_\dof & \\
      & & & & - \lambda_\dof
    \end{pmatrix} 
    \begin{pmatrix}
      x_1 \\
      x_2 \\
      \vdots \\
      x_{2\dof-1} \\
      x_{2\dof}
    \end{pmatrix} \,.
  \label{eq:NF-diagonal-matrix-A}
\end{equation}%
Note that the originally real vector field \eqref{eq:NF-DGL-starting} is, in
general, transformed into a complex one ($\aa_n \in \Cbbm^{2\dof}$) by the
diagonalisation, if the eigenvalue spectrum includes imaginary terms.
Analogously to the dynamical equations, also the energy functional is expanded
in the vicinity of the fixed point, and the linear transformation $\xx \to \xx'
= T^{-1} \xx$ is applied. This results in the scalar field
\begin{equation}
  E(\xx) = \sum_{n=0}^{\nmax+1} E_n(\xx) \,, 
  \label{eq:NF-energy-func-starting}
\end{equation}
whose coefficients, again, become complex in general. The zeroth-order term 
$E_0$ is the fixed-point energy, the first-order of the expansion vanishes, 
$E_1=0$, and, with the normalization \eqref{eq:NF-eigenvectors-normalization}, 
the second-order term has the structure $E_2 = \sum_{i=1}^\dof \lambda_i \, 
x_{2i-1} x_{2i}$. Consequently, the energy functional 
\eqref{eq:NF-energy-func-starting} is in Poincar\'e-Birkhoff normal form up to 
the order $n=2$.

\subsection{Normal form transformations -- nonresonant terms}
\label{sec:NF-step2}

The diagonalisation of the local dynamical equations as described in the
previous section \ref{sec:NF-step1}, simplifies their linear part in a way that 
it becomes 
diagonal.
However, for the terms of higher order, a ``simplification'' cannot be achieved 
by this step. For this purpose, a normal form expansion of the diagonalized 
dynamical equations \eqref{eq:NF-DGL-starting-diagonalized-aa} is performed in 
this section making use of successive Lie transforms.
The general treatment of local dynamical systems and their normal forms has been
described by Murdock \cite{Murdock2010} in detail.
Here, it is applied to the $2\dof$-dimensional local dynamical equations 
\eqref{eq:NF-DGL-starting-diagonalized-aa} with their special eigenvalue 
structure \eqref{eq:TDVP-EV-structure-pm-structure}.

In order to bring the local equations of motion into normal form, a nonlinear
near-identity transformation 
\begin{equation}
  \xx = \vec{\phi}_\eps (\yy) 
   \label{eq:NF-change-of-coordinates}
\end{equation}
is applied, which transforms from the ``old'' coordinates $\xx$ to ``new'' ones
$\yy$, and which is differentiable in the new coordinates $\yy$ as well as in
the parameter $\eps$. The latter serves as a continuous scaling parameter that
is introduced in a way that for $\eps=0$ one obtains the identity
transformation, while the finally desired transformation is obtained for
$\eps=1$,
\begin{subequations}%
\begin{align}
   \xx &= \vec{\phi}_{\eps=0}(\yy) = \yy \,,  
   \label{eq:NF-identity-transformation} \\
   \xx &= \vec{\phi}_{\eps=1}(\yy) = \vec{\phi}(\yy) \,.  
   \label{eq:NF-full-transformation}
\end{align}%
\end{subequations}
Instead of providing the explicit function \eqref{eq:NF-change-of-coordinates},
the change of coordinates is defined implicitly, by the requirement that it is
the solution of the differential equation
\begin{equation}
   \frac{\dd \xx}{\dd \eps} = \vec\generator (\xx) \,, 
   \label{eq:NF-generating-function-definition}
\end{equation}
with $\vec\generator$ being the generating function of the transformation.
As it is shown in \REF\ \cite{Murdock2010}, the final change of 
variables \eqref{eq:NF-full-transformation} transforms a vector field $\aa$ 
defining the differential equation
\begin{equation}
   \dod{\xx}{t} = \aa(\xx) 
   \label{eq:NF-vector-field-starting}
\end{equation}
into a vector field $\bb$ in the new coordinates $\yy$ with
\begin{equation}
   \dod{\yy}{t} = \bb(\yy) \,. 
   \label{eq:NF-vector-field-transformed}
\end{equation}
The connection between the two vector fields is
\begin{equation}
   \vec{b} (\yy)
   =
   \sum_{j=0}^\infty
   \frac{1}{j!}
   \strich{\Lop{\vec\generator}{j}
   \vec a (\xx)}_{\xx=\yy} \,, 
   \label{eq:NF-time-one-map-trafo-vector}
\end{equation}
where $\Lop{\vec\generator}{}$ is the \emph{homological operator} acting on
differentiable vector fields according to
\EDITS{
\begin{equation}
   \Lop{\vec\generator}{} a_m (\xx)
   \defined \sum_{n=1}^{2\dof}
   \pop{{a}_m(\xx)}{x_n} \, {g}_n(\xx) -
   \pop{{g}_m(\xx)}{x_n} \, {a}_n (\xx) \,. 
  \label{eq:NF-definition-Lie-operator}
\end{equation}
}
Analogously, the same generating function transforms the energy functional
according to
\begin{align}
   \tilde E (\yy)
   &= \sum_{j=0}^\infty
      \frac{1}{j!} \strich{
      \Dop{\vec\generator}{j}
              E (\xx)}_{\xx=\yy} \,. 
   \label{eq:NF-time-one-map-trafo-scalar} 
\end{align}
Here, the \emph{right-multiplication operator} $\Dop{\vec{g}}{}$ is defined by
\begin{equation}
   \Dop{\vec{\generator}}{} E(\xx)
   \definition
   \pop{E(\xx)}{\xx} \,
   \vec{\generator} (\xx) \,. 
\label{eq:NF-definition-D-operator}
\end{equation}

\subsubsection{Transformation of multivariate polynomials}

As already mentioned above, the local dynamical equations as well as the
energy functional are on hand in the form of a formal power series or local
Taylor expansion, \ie\ as a multivariate polynomial. Therefore, it is convenient
to also define the generating function as a multivariate polynomial, so that the
transformed fields will also be such ones.
In the following, these polynomials are written as
\begin{subequations}
\begin{align}
   \aa (\xx)
   &= \sum_{n=1}^\nmax \aa_n (\xx)
    = \sum_{\abs{\mm}=1}^\nmax
      \vec\alpha_{\mm} \, \xx^\mm \,,  
\label{eq:NF-multivariate-expansion-a} \\
   \bb (\xx)
   &= \sum_{n=1}^\nmax \bb_n (\xx)
    = \sum_{\abs{\mm}=1}^\nmax
      \vec\beta_{\mm} \, \xx^\mm \,,  \\
   \vec\generator (\xx)
   &= \sum_{n=1}^\nmax \vec\generator_n (\xx)
    = \sum_{\abs{\mm}=1}^\nmax
      \vec\gamma_{\mm} \, \xx^\mm \,,  \\
  E(\xx) 
  &=
  \sum_{n=0}^{\nmax+1} E_n(\xx)
  = \sum_{\abs{\mm}=0}^{\nmax+1} 
    \xi_{\mm} \, \xx^{\mm} \,, 
\label{eq:NF-multivariate-expansion-d} 
\end{align}%
\label{eq:NF-multivariate-expansion}%
\end{subequations}
where $\aa_n, \bb_n, \vec \generator_n, E_n$ denote the terms of the respective 
series which are homogeneous of degree $n$, and $\vec\alpha_{\mm}, 
\vec\beta_{\mm}, \vec\gamma_{\mm}, \xi_{\mm}$ are the coefficients of the 
expansion.
Furthermore, the multi-index 
notation
\begin{subequations}
\begin{align}
  \xx^\mm   &=
  x_1^{m_1} \, x_2^{m_2} \, \ldots \, x_{2\dof}^{m_{2\dof}} \,,  \\
  \abs{\mm} &=
  m_1 + m_2 + \ldots + m_{2\dof} 
\end{align}%
\label{eq:NF-multiindex-notation}%
\end{subequations}
with the integer vector $\mm \in \Nbbm^{2\dof}_0$ is used.
The purpose of the following normal form transformation is that -- for given
expansion coefficients $\alpha_{\mm k}$ and $\xi_{\mm k}$ -- the coefficients
$\gamma_{\mm k}$ of the generating function are chosen in such a way that as 
many as possible of the resulting coefficients $\beta_{\mm k}$ vanish, and that
they are connected to the energy functional via canonical equations.

\begin{mydef}
\EDITS{
Define the set $\Mset$ of integer vectors by
\begin{equation}
  \Mset \defined \left\lbrace
    \mm \in \Nbbm^{2\dof} \, \bigl. \bigr| \,
    m_{2j-1} = m_{2j} \, ; \,
    j=1,2,\ldots,\dof
  \right\rbrace \,.
  \label{eq:NF-Mset}
\end{equation}
}
\end{mydef}
\EDITS{
Then the transformed dynamical equation take the following form.
}
\begin{thm}[Polynomial structure of the transformed dynamical equations]
\EDITS{
If the eigenvalues $\lambda_i$ of the linearized dynamical equations are 
pair-wise rationally independent (\ie~$\lambda_i/\lambda_j \notin \Qbbm$, of all 
pairs of eigenvalues $i \neq j$), an appropriate generating function 
$\vec\generator$ transforms the vector field $\aa$ into the general polynomial 
structure (\cf\ Table \ref{tab:monomial-structure-example})
%
\begin{subequations}
\begin{align}
  b_{n (2i-1)}
  &=
  \sum_{\substack{\mm \in \Mset, \\ \abs{\mm}=n+1, \\ \nonneg}}
  \beta_{[\mm - \unitvec_{2i}] (2i-1)} \,
  \xx^{[\mm - \unitvec_{2i}]} \,, 
\\
  b_{n (2i)}
  &=
  \sum_{\substack{\mm \in \Mset, \\ \abs{\mm}=n+1, \\ \nonneg}}
  \beta_{[\mm - \unitvec_{2i-1}] (2i)} \,
  \xx^{[\mm - \unitvec_{2i-1}]} \,. 
\end{align}
\label{eq:NF-polynomial-structure-DGLs}%
\end{subequations}
with $i=1,2, \ldots, \dof$.
Here, the summation is carried out over the set \eqref{eq:NF-Mset} and
$\unitvec_{2i-1}$ as well as $\unitvec_{2i}$ are unit vectors.
%
Moreover, the constraint ``$\nonneg$'' in the summation denotes to add only
those terms for which the indices $[\mm - \unitvec_{2i}]$ and $[\mm - 
\unitvec_{2i-1}]$ only have nonnegative entries, \ie\ those $\mm \in \Mset$
with 
$m_{2i-1}=m_{2i}=0$ are not taken into account.
An equivalent interpretation of this constraint is to set all terms $\beta_{\mm
k}$ to zero, if its index $\mm$ possesses at least one negative entry.
}
\end{thm}

\begin{proof}
\EDITS{
One can specifically transform the $n$-th order of the original vector field 
$\vec a_n$, if the generating function is chosen to be homogeneous of degree 
$n$,
\begin{equation}
  \vec\generator (\xx) =
  \vec\generator_n (\xx) =
  \sum_{\abs{\mm}=n} \vec\gamma_\mm \, \xx^\mm \,. 
  \label{eq:NF-generator-degree-i}
\end{equation}
Inserting the multivariate polynomials \eqref{eq:NF-multivariate-expansion} with
the constraint \eqref{eq:NF-generator-degree-i} for the generating function into
\EQ\ \eqref{eq:NF-time-one-map-trafo-vector}, one obtains, after renaming $\yy$
by $\xx$, the homological equation
\begin{align}
\begin{split}
   \vec b_n (\xx)
   &=
   \vec a_n(\xx)
   +
   \Lop{\vec\generator_n}{} \, \vec a_1(\xx)   
   \label{eq:NF-homological-equation}
\end{split}
\end{align}
for the transformation of the monomials which are equal to the degree of the
generating function.
From this, the transformation of the single coefficients can be extracted. For
the $k$-th component ($k=1,2,\ldots, 2\dof$) it reads
\begin{equation}
   \beta_{\mm k}
   =
   \alpha_{\mm k} +
   \bigl( \lambda_k^\pm - \left< \mm, \vec\lambda^\pm \right> \bigr) \,
   \gamma_{\mm k} \,. 
   \label{eq:NF-trafo-coefficients}
\end{equation}
where $\left< \mm, \vec\lambda^\pm \right>$ is the standard scalar product.
%
One can see from \EQ\ \eqref{eq:NF-trafo-coefficients} that a nonvanishing
monomial ($\alpha_{\mm k} \neq 0$) can be eliminated ($\beta_{\mm k} = 0$) by
the Lie transform with an appropriate choice of the generating function, if the
eigenvalue $\lambda_k^\pm$ is ``nonresonant'', \ie\ if
\begin{subequations}%
\begin{align}
   \lambda_k^\pm - \left< \mm, \vec\lambda^\pm \right> &\neq 0 \,. 
   \label{eq:NF-cond-resonance-false} \\
\intertext{Otherwise, if the condition of resonance}
   \lambda_k^\pm - \left< \mm, \vec\lambda^\pm \right> &= 0 
   \label{eq:NF-cond-resonance-true}
\end{align}%
\label{eq:NF-cond-resonance-both}%
\end{subequations}
is fulfilled, the respective term cannot be eliminated. 
The final polynomial structure of the normal form of the local dynamical
equations is determined by the eigenvalues of the linearised equations of
motion, because only monomials fulfilling \EQ\ \eqref{eq:NF-cond-resonance-true}
remain after the Lie transforms. Moreover, due to the fact that these
eigenvalues exhibit the general structure
\eqref{eq:TDVP-EV-structure-pm-structure} of pairwise eigenvalues with different
sign, the normal form also possesses a general polynomial structure.
Denoting the entries of the integer vector by $ \mm = ( m_1, m_2, \ldots ,
m_{2\dof} )^\transpose$, the condition of resonance 
\eqref{eq:NF-cond-resonance-true} becomes ($i=1,2,\ldots,\dof$)
\begin{equation}
  \bigl[
  \lambda_1    (m_1    - m_2   ) +
  \lambda_2    (m_3    - m_4   ) +
  \ldots +
  \lambda_\dof (m_{2\dof-1} - m_{2\dof})
  \bigr]
  =
  \pm \lambda_i \,,
  \label{eq:NF-cond-resonance-true-splitted}
\end{equation}
where the upper sign is valid for $k=2i-1$ and the lower one for $k=2i$.
Assuming pair-wise rational independence of the eigenvalues, 
\EQ~\eqref{eq:NF-cond-resonance-true-splitted} is fulfilled if and only if
\begin{subequations}
\begin{align}
  && m_{2i-1} &= m_{2i} \pm 1 \,, 
  && (i=1,2,\ldots,\dof) \,, &&
  \label{eq:NF-remaining-monomials-mm-differ} \\
  && m_{2j-1} &= m_{2j}     \,, 
  &&  (j \neq i) \,. &&
  \label{eq:NF-remaining-monomials-mm-same}%
\end{align}
\label{eq:NF-remaining-monomials-mm}%
\end{subequations}
}
\end{proof}

\begin{mydef}
\EDITS{
In the following, monomials $\xx^\mm$ whose integer vector $\mm$ fulfils \EQ\ 
\eqref{eq:NF-cond-resonance-false} are referred to as ``nonresonant monomials'' 
and those fulfilling \EQ~\eqref{eq:NF-cond-resonance-true} are called 
``resonant monomials''. 
Analogously, their coefficients are referred to as nonresonant and resonant 
coefficients, respectively.
}
\end{mydef}
\begin{mydef}
\EDITS{
The vector field $\aa$ is said to be in normal form with respect to its linear 
part $\vec a_1$, if it only contains monomials fulfilling 
\EQ~\eqref{eq:NF-cond-resonance-true}.
}
\end{mydef}

%
\begin{table}[t]
\caption[
Illustration of the fundamental polynomial structure of the dynamical equations
and the energy functional in normal form coordinates
]{%
Illustration of the fundamental polynomial structure of the dynamical equations
\eqref{eq:NF-polynomial-structure-DGLs} and the energy functional
\eqref{eq:NF-polynomial-structure-E} for a system with $\dof = 2$ degrees of
freedom.
In normal form coordinates, there remain only terms with odd degree of the
monomial in the equations of motion. Moreover, the exponents of the variables
$x_{2i-1}, x_{2i}$ in the respective component of the vector field differ by one
and the terms $x_{2j-1}, x_{2j}$ with $j \neq i$ (displayed in brackets) have
the same exponent.
By contrast, the energy functional only consists of monomials with even degree
and all variables $x_{2j-1}, x_{2j}$ occur in products.
The extension of this structure to $\dof>2$ degrees of freedom is
straightforward. In this case, additional terms $(x_5 x_6), (x_7 x_8), \ldots$
occur in the expansions.
}
\vspace{1em}
\centering
\renewcommand{\arraystretch}{1.3}
\scriptsize
\begin{tabular}{cccccccc}
  \toprule
  ~ & \multicolumn{7}{c}{Degree $n$ of the monomial}
\\ \cmidrule{2-8}
  Field &
  0 & 1 & 2 & 3 & 4 & 5 & 6
\\ \midrule
  $\dot x_1$ & -- &
  $x_1$ &
  -- &
  $x_1^2 x_2^1 (x_3 x_4)^0$ &
  -- &
  $x_1^3 x_2^2 (x_3 x_4)^0$ &
  --
\\
  & & & &
  $x_1^1 x_2^0 (x_3 x_4)^1$ & &
  $x_1^2 x_2^1 (x_3 x_4)^1$ &
\\
  & & & & & &
  $x_1^1 x_2^0 (x_3 x_4)^2$ &
\\ \cmidrule{2-8}
  $\dot x_2$ & -- &
  $x_2$ &
  -- &
  $x_1^1 x_2^2 (x_3 x_4)^0$ &
  -- &
  $x_1^2 x_2^3 (x_3 x_4)^0$ &
  --
\\
  & & & &
  $x_1^0 x_2^1 (x_3 x_4)^1$ & &
  $x_1^1 x_2^2 (x_3 x_4)^1$ &
\\
  & & & & & &
  $x_1^0 x_2^1 (x_3 x_4)^2$ &
\\ \midrule
  $\dot x_3$ & -- &
  $x_3$ &
  -- &
  $x_3^2 x_4^1 (x_1 x_2)^0$ &
  -- &
  $x_3^3 x_4^2 (x_1 x_2)^0$ &
  --
\\
  & & & &
  $x_3^1 x_4^0 (x_1 x_2)^1$ & &
  $x_3^2 x_4^1 (x_1 x_2)^1$ &
\\
  & & & & & &
  $x_3^1 x_4^0 (x_1 x_2)^2$ &
\\ \cmidrule{2-8}
  $\dot x_4$ & -- &
  $x_4$ &
  -- &
  $x_3^1 x_4^2 (x_1 x_2)^0$ &
  -- &
  $x_3^2 x_4^3 (x_1 x_2)^0$ &
  --
\\
  & & & &
  $x_3^0 x_4^1 (x_1 x_2)^1$ & &
  $x_3^1 x_4^2 (x_1 x_2)^1$ &
\\
  & & & & & &
  $x_3^0 x_4^1 (x_1 x_2)^2$ &
\\ \midrule
  $E$ &
  const. &
  -- &
  $(x_1 x_2)^1$ &
  -- &
  $(x_1 x_2)^2 (x_3 x_4)^0$ &
  -- &
  $(x_1 x_2)^3 (x_3 x_4)^0$
\\
  & & &
  $(x_3 x_4)^1$ & &
  $(x_1 x_2)^1 (x_3 x_4)^1$ & &
  $(x_1 x_2)^2 (x_3 x_4)^1$
\\
  & & & & &
  $(x_1 x_2)^0 (x_3 x_4)^2$ & &
  $(x_1 x_2)^1 (x_3 x_4)^2$
\\
  & & & & & & & $(x_1 x_2)^0 (x_3 x_4)^3$
\\ \bottomrule
\end{tabular}
\label{tab:monomial-structure-example}
\end{table}
%

Concluding, in normal form coordinates, the variables $x_{2i-1}, x_{2i}$ in the
respective component of the dynamical equations occur with exponents which
differ exactly by one, while the terms $x_{2j-1}, x_{2j}$ with $j \neq i$ have
the same exponent (\cf\ Table \ref{tab:monomial-structure-example}).
Note that all monomials remaining in \EQS\
\eqref{eq:NF-polynomial-structure-DGLs} are of odd degree. All terms of even
degree have been eliminated completely by the Lie transforms, because the
condition of resonance \eqref{eq:NF-cond-resonance-true} cannot be fulfilled, if
$\abs{\mm}$ is even.

\subsubsection{Determination of the generating function to eliminate the 
nonresonant terms}

After having discussed the general structure of the normal form, its actual 
calculation is presented in this section. The calculation will be carried out 
order by order, \ie\ the orders $n=2,3,4,\ldots$ are treated successively. 
It is assumed that the system is already in normal form up to the order $n-1$. 
Then, a generating function $\vec\generator_n$ is constructed to transform the 
$n$-th order of the equations of motion. 

As already mentioned above, the coefficients $\gamma_{\mm k}$ of the generating 
function which are nonresonant, \ie\ \EQ\ \eqref{eq:NF-cond-resonance-false} is 
valid, can be chosen in such a way that the corresponding term $\alpha_{\mm k}$ 
of the original expansion is eliminated.
Such nonresonant coefficients occur in every order of the expansion. In
particular, the generating function of each even degree $n$ only consists of
nonresonant coefficients.
The determination of the nonresonant coefficients $\gamma_{\mm k}$ of the
generating function is straightforward. Since their purpose is to eliminate the
original term $\alpha_{\mm k}$, they are uniquely determined by \EQ\
\eqref{eq:NF-trafo-coefficients}. Requiring $\beta_{\mm k}=0$ and solving for
the coefficient of the generating function \eqref{eq:NF-generator-degree-i}, one
obtains
\begin{equation}
   \gamma_{\mm k} =
   \begin{cases}
      \cfrac{\alpha_{\mm k}}
            {\left< \mm, \vec\lambda^\pm \right> - \lambda_k^\pm} \,, \qquad
      &\text{if } \lambda_k^\pm - \left< \mm, \vec\lambda^\pm \right> \neq 0 
\,, \\[1em]
      c_{\mm k} \,, &\text{else}.
   \end{cases} 
\label{eq:NF-coefficients-generator-construction}
\end{equation}
The choice in the first line guarantees the elimination of the term $\alpha_{\mm
k}$ in the nonresonant case. All coefficients $c_{\mm k}$ related to the 
resonant terms are free parameters, which do not change the $\beta_{\mm k}$ of 
the order $\abs{\mm}=n$. For simplicity these terms are set to $c_{\mm k}=0$ in 
the transformations of the nonresonant coefficients and their final 
determination will be treated separately (see \SEC\ \ref{sec:NF-step3}).

\begin{thm}[Polynomial structure of the transformed energy functional]
\EDITS{
The application of the generating function $\vec\generator$ with coefficients 
fulfilling \EQ~\eqref{eq:NF-coefficients-generator-construction} to the energy 
functional, transforms the latter into the general polynomial structure (\cf\ 
Table \ref{tab:monomial-structure-example})
\begin{equation}
  E_{n+1}(\xx) =
  \sum_{\substack{\mm \in \Mset \\ \abs{\mm}=n+1}}
  \xi_{\mm} \, \xx^{\mm} \,. 
  \label{eq:NF-polynomial-structure-E}
\end{equation}
}
\end{thm}
\begin{proof}
\EDITS{
As already mentioned in \SEC~\ref{sec:TDVP}, the symplectic 2-form 
\eqref{eq:TDVP-2-form-K-tilde} is skew-symmetric, nondegenerate, and closed.
Therefore, Darboux' theorem guarantees the existence of canonical coordinates 
fulfilling the relation \eqref{eq:Ham-Eqs-gamma-Jlinks}.
Any polynomial structure differing from \EQ~\eqref{eq:NF-polynomial-structure-E} 
would result in terms that have no relation in the corresponding 
dynamical equations \eqref{eq:NF-polynomial-structure-DGLs} and would, 
therefore, violate Darboux' theorem.
}
\end{proof}

\subsection{Normal form transformations -- resonant terms}
\label{sec:NF-step3}

In normal form coordinates the dynamical equations
\eqref{eq:NF-polynomial-structure-DGLs} and the energy functional
\eqref{eq:NF-polynomial-structure-E} naturally exhibit a polynomial structure
that allows for the identification of the normal form coordinates as canonical
ones according to the canonical equation
\begin{align}
  \bb_n (\xx) = \symplecticJ \, \pop{}{\xx} \, E_{n+1} (\xx)
\label{eq:NF-canonequations-for-NF}%
\end{align}
with the energy functional acting as Hamiltonian. Equation
\eqref{eq:NF-canonequations-for-NF} is valid in each order $n$, if the
coefficients $\beta_{\mm k}$ and $\xi_{\mm}$ in \EQS\
\eqref{eq:NF-polynomial-structure-DGLs} and \eqref{eq:NF-polynomial-structure-E}
fulfil the conditions
\begin{subequations}
\begin{align}
  m_{2i} \xi_{\mm} &=
  \beta_{[\mm - \unitvec_{2i}] (2i-1)} \,, 
  \label{eq:NF-integration-conditions-a}  \\
  \beta_{[\mm-\unitvec_{2i}] (2i-1)} &=
 -\beta_{[\mm-\unitvec_{2i-1}] (2i)} \,, 
  \label{eq:NF-integration-conditions-b}  \\
  \frac{\beta_{[\mm-\unitvec_{2i}] (2i-1)}}{m_{2i}} &=
  \frac{\beta_{[\mm-\unitvec_{2j}] (2j-1)}}{m_{2j}} 
  \label{eq:NF-integration-conditions-c}  
\end{align}
\label{eq:NF-integration-conditions-all}%
\end{subequations}
for all $i,j = 1,2,\ldots,\dof$ ($i\neq j$) and $\mm \in \Mset$ with
$\abs{\mm}=n+1$.
Here, \EQ\ \eqref{eq:NF-integration-conditions-a} is the requirement that the 
coefficients of the dynamical equations and those of the energy functional are
connected via derivatives according to \EQS\
\eqref{eq:NF-canonequations-for-NF}.
The sign structure of the symplectic matrix $\symplecticJ$ is taken into account
by \EQ\ \eqref{eq:NF-integration-conditions-b} for each pair $x_{2i-1}, x_{2i}$
of ``conjugate coordinates'',
and \EQ\ \eqref{eq:NF-integration-conditions-c} considers the fact that terms in
the expansion of \emph{different} pairs $x_{2i-1},x_{2i}$ and $x_{2j-1}, x_{2j}$
($i\neq j$) result from the \emph{same} term of the energy functional.

As a consequence of the normal form expansion together with the general 
eigenvalue structure \eqref{eq:TDVP-EV-structure-pm-structure}, \EQ\ 
\eqref{eq:NF-integration-conditions-b} is fulfilled after the Lie transforms 
have been applied as discussed in \SEC\ \ref{sec:NF-step2}. However, the 
conditions \eqref{eq:NF-integration-conditions-a} and 
\eqref{eq:NF-integration-conditions-c} are \emph{not} fulfilled, in general.
This is due to the fact that -- although the polynomial structure of the
expansions is uniquely determined by the eigenvalue structure -- the explicit
normal form, \ie\ the coefficients of the expansion, are not unique. The reason
is that the resonant coefficients $c_{\mm k}$ of the generating function in \EQ\
\eqref{eq:NF-coefficients-generator-construction} are free, and that the choice
to set them zero does not guarantee the fulfilment of all \EQS\
\eqref{eq:NF-integration-conditions-all}. 
Therefore, further steps are necessary in order to guarantee that the latter are
valid, and these steps are presented in the following.

We emphasize that it is precisely this treatment of the resonant terms of the 
generating function which is the difference between the usual normal form 
procedure of canonical Hamiltonians and the transformation of the noncanonical 
system: If the coordinates had been canonical at the beginning, the choice 
$c_{\mm k}=0$ in \EQ\ \eqref{eq:NF-coefficients-generator-construction} would 
have kept this property. Vice versa, we will use an appropriate choice $c_{\mm 
k}\neq 0$ in the following to generate canonical coordinates.

\subsubsection{Particular choice of the resonant generating function and the
corresponding transformations}

Resonant terms occur in every odd order $n=3,5,7,\ldots$ of the generating
function \eqref{eq:NF-generator-degree-i}, and a fundamental property of them is
the fact that they do not affect the polynomial structure of the expansions, but
they only modify their coefficients. 
Vice versa, this property can be used in order to guarantee the fulfilment of
the canonical equations \eqref{eq:NF-integration-conditions-all} by a suitable
choice of the resonant terms as it will be demonstrated in the following. 
For this purpose, it is investigated in detail in this section how a resonant
generating function of degree $n$ transforms the next-higher order terms of the
dynamical equations as well as the energy functional.
Finally, \EQS\ \eqref{eq:NF-integration-conditions-all} will serve as
conditional equations for the determination of the resonant coefficients.

In a resonant generating function of degree $n$, there occur coefficients
$\gamma_{[\mm-\unitvec_{2i}](2i-1)}$ and $\gamma_{[\mm-\unitvec_{2i-1}](2i)}$
with $\mm \in \Mset$ and $\abs{\mm}=n+1$, \ie\ there is exactly one term 
corresponding to each of the monomials remaining in the dynamical equations 
\eqref{eq:NF-polynomial-structure-DGLs}.
In order to guarantee that \EQS\ \eqref{eq:NF-integration-conditions-all} hold 
for the whole expansion, it will be sufficient only to consider the terms 
$\gamma_{[\mm-\unitvec_{2i}](2i-1)}$ and to set 
$\gamma_{[\mm-\unitvec_{2i-1}](2i)} = 0$ for simplicity. 
With this choice, the resonant generating function homogeneous of degree $n$ has
the form
\begin{subequations}
\begin{align}
  \generator_{n (2i-1)} (\xx)
  &=
  \sum_{\substack{\mm \in \Mset, \\ \abs{\mm} = n+1, \\ \nonneg}}
  \gamma_{[\mm - \unitvec_{2i}] (2i-1)} \,
  \xx^{[\mm - \unitvec_{2i}]} \,, 
\\
  \generator_{n (2i)} (\xx)
  &=
  0 \,.
\end{align}
\label{eq:NF-res-g-choice}%
\end{subequations}
It is easily verified from \EQ\ \eqref{eq:NF-definition-D-operator} that the
$j$-fold application $\Dop{\vec\generator_n}{j} E_k$ of the right-multiplication
operator with a generating function homogeneous of degree $n$ onto the part of
the energy functional of degree $k$ results in a polynomial homogeneous of
degree $k+j(n-1)$. Furthermore, the lowest order which is affected in \EQ\
\eqref{eq:NF-time-one-map-trafo-scalar} is $n+1$.
Consequently, there are two cases which contribute to the order $n+1$ of the
transformed field, namely those with $k+j(n-1) \stackrel{!}{=} n+1$. On the one
hand, this is the contribution $k=n+1$ and $j=0$, on the other hand it is $k=2$
and $j=1$, so that the precise transformation reads
\begin{align}
  \tilde E_{n+1} (\xx)
  =
  E_{n+1} (\xx)
  + \Dop{\vec\generator_n}{} E_2 (\xx) \,,
\label{eq:NF-res-E-trafo-gn}
\end{align}
where $\tilde E_{n+1} = \sum_{\abs{\mm}=n+1} \tilde \xi_\mm \, \xx^\mm$ is the
transformed field.
Inserting the expansions \eqref{eq:NF-multivariate-expansion} into \EQ\
\eqref{eq:NF-res-E-trafo-gn}, and using the fact that the second order of the
energy functional has the form $E_2 = \sum_{i=1}^\dof \lambda_i\, 
x_{2i-1}\, x_{2i}$ in normal form coordinates, \EQ\ 
\eqref{eq:NF-res-E-trafo-gn} can directly be reformulated in terms of the 
energy functional's coefficients:
\begin{equation}
  \tilde \xi_\mm
  = 
  \xi_\mm +
  \sum_{\substack{i=1, \\ \nonneg}}^\dof \lambda_i \,
  \gamma_{[\mm-\unitvec_{2i}](2i-1)} \,.
  \label{eq:NF-res-trafo-E-gn-coeff}
\end{equation}
Analogously, the effect of a resonant generating function onto the dynamical
equations can be investigated.
If the latter are already in their normal form
\eqref{eq:NF-polynomial-structure-DGLs}, there are only odd degrees of the
expansion $\bb_1, \bb_3, \bb_5, \ldots$ left as discussed above. 
By definition, the first-order term $\bb_1$ containing the eigenvalues does not 
contribute to the Lie operator for a resonant generating function, \ie\ 
$\Lop{\vec\generator_n}{}\bb_1 = 0$. This identity directly follows from \EQ\ 
\eqref{eq:NF-trafo-coefficients}, because the resonant coefficients are always 
multiplied by zero. Therefore, the lowest-order term which leads to a 
contribution of the Lie operator is the term $\bb_3$.
From \EQ\ \eqref{eq:NF-time-one-map-trafo-vector} it follows that the
lowest-order term which is modified by a resonant generating function of degree
$n$ together with $\bb_3$ is the order $n+2$ of the dynamical equations, 
\begin{equation}
  \tilde \bb_{n+2} (\xx) 
  = 
  \bb_{n+2} (\xx) 
  + \Lop{\vec\generator_n}{} \bb_3(\xx) \,.
\label{eq:NF-res-trafo-gn-DGLs}
\end{equation}
Multiple applications of the Lie operator as well as higher-order terms $\bb_n$
with $n>3$ lead to higher-order corrections and do not need to be considered
here.
Analogously to the energy functional, this transformation can be rewritten 
directly in terms of the vector field's coefficients. After a short 
calculation, one obtains 
\begin{subequations}
\begin{align}
  \tilde \beta_{[\mm-\unitvec_{2i}](2i-1)}
  &=
    \beta_{[\mm-\unitvec_{2i}](2i-1)} 
  + \summea \,,
\\[.1em]
  \tilde \beta_{[\mm-\unitvec_{2i-1}](2i)}
  &=
    \beta_{[\mm-\unitvec_{2i-1}](2i)} 
  + \summec 
\end{align}
\label{eq:NF-res-trafo-gn-b3-coeff}%
\end{subequations}
with the quantities
\begin{subequations}
\begin{align}
\begin{split}
  \summea 
  &\defined
  \sum_{\substack{\mm'' \in \Mset, \\ \abs{\mm''}=4, \\ \nonneg}}
  \Biggl[
    - (m''_{2i}-1) \,
    \beta_{[\mm''-\unitvec_{2i-1}](2i)} \,
    \gamma_{[\mm - \mm'' + \unitvec_{2i-1}](2i-1)} 
  \Biggr.
\\[-2em]
  &\qquad\qquad\quad
  + \sum_{\substack{i'=1\\i'\neq i}}^\dof m''_{2i'} \,
  \bigl(
    \beta_{[\mm'' - \unitvec_{2i}](2i-1)} \,
    \gamma_{[\mm - \mm'' + \unitvec_{2i'-1}](2i'-1)}
  \bigr.
\\[-1em] & \qquad\qquad\qquad \phantom{+\sum} \bigl.
    - \beta_{[\mm'' - \unitvec_{2i'}](2i'-1)} \,
    \gamma_{[\mm - \mm'' 
      + \unitvec_{2i'-1} + \unitvec_{2i'} - \unitvec_{2i}](2i-1)} 
\\ & \qquad\qquad\qquad \phantom{+\sum} 
    - \beta_{[\mm'' - \unitvec_{2i'-1}](2i')} \,
    \gamma_{[\mm - \mm'' 
      + \unitvec_{2i'-1} + \unitvec_{2i'} - \unitvec_{2i}](2i-1)}
  \bigr) \Biggr] \,,
\end{split}
\allowdisplaybreaks[4]
\\ 
\begin{split}
  \summec 
  &\defined
  \sum_{\substack{\mm'' \in \Mset, \\ \abs{\mm''}=4, \\ \nonneg}}
  \Biggl[
    (m''_{2i}-1) \,
    \beta_{[\mm''-\unitvec_{2i-1}](2i)} \,
    \gamma_{[\mm - \mm'' + \unitvec_{2i-1}](2i-1)} 
  \Biggr.
\\[-2em]
  & \qquad\qquad\qquad\qquad  \Biggl. +
  \sum_{\substack{i'=1\\i'\neq i}}^\dof
  m''_{2i'} \,
  \beta_{[\mm'' - \unitvec_{2i-1}](2i)} \,
  \gamma_{[\mm - \mm'' + \unitvec_{2i'-1}](2i'-1)} 
  \Biggr] \,.
\end{split}
\end{align}
\label{eq:NF-res-def-sigmas}%
\end{subequations}
It is emphasized that $\summea$ and $\summec$ depend on the resonant 
coefficients $\gamma_{\mm k}$ \emph{linearly} and that only the third-order 
coefficients $\beta_{\mm k}$ occur therein.

\subsubsection{Determination of the generating function's resonant coefficients}

As already mentioned above, the resonant coefficients of each generating
function are free parameters, in the sense that they do neither change the
polynomial structure of the dynamical equations nor that of the energy
functional. However, they do modify the coefficients of the respective 
expansions.
\begin{thm}[Choice of the resonant generating function]
\EDITS{
The choice of the resonant coefficients of the generating function according to
\begin{subequations}
\begin{align}
  \gamma_{[\mm-\unitvec_{2i-1}](2i)}&=0 \\
  \sum_{\substack{i=1 \\ \nonneg}}^\dof \lambda_i 
  \gamma_{[\mm-\unitvec_{2i}](2i-1)}
  &=
  \tilde \xi_{\mm} - \xi_{\mm}\,,
  \label{eq:NF-res-LGS-a} 
\\
  m_{2j} \, \summea
  -
  m_{2i} \, \summeaj 
  &= 
  m_{2i} \, \beta_{[\mm-\unitvec_{2j}](2j-1)} -
  m_{2j} \, \beta_{[\mm-\unitvec_{2i}](2i-1)} 
\label{eq:NF-res-LGS-b}
\end{align}
\label{eq:NF-res-LGS}%
\end{subequations}
guarantees that all the conditions of integration 
\eqref{eq:NF-integration-conditions-all} are fulfilled, \ie~the final set of 
coordinates are standard canonical coordinates.
}
\end{thm}


%
\begin{figure}[t]
\centering
\flushright
\includegraphics[width=.8\columnwidth]{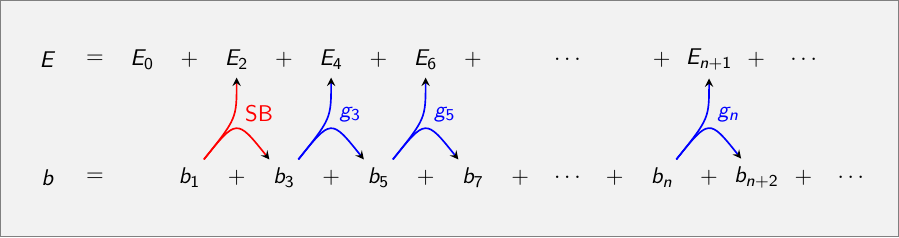}
\vspace{1em}
\caption{%
Scheme of the procedure to determine the resonant coefficients of the
generating function.
In each order $n \geq 3$, it is assumed that the term $\bb_n$ of the dynamical 
equations fulfils \EQS\ \eqref{eq:NF-integration-conditions-b} and 
\eqref{eq:NF-integration-conditions-c}.
Then, the resonant coefficients of the generating function are determined using 
the part $E_{n+1}$ of the energy functional and the next-higher order term 
$\bb_{n+2}$ of the equations of motion (blue). 
The resonant generating function is constructed in a way that $\bb_n$ as well
as 
$E_{n+1}$ are connected via the canonical equation 
\eqref{eq:NF-canonequations-for-NF} and that \EQS\ 
\eqref{eq:NF-integration-conditions-b} and 
\eqref{eq:NF-integration-conditions-c} are valid for the term $\bb_{n+2}$.
The whole procedure is applied successively for the orders $n=3,5,7,\ldots$
Note that the same conditions are fulfilled automatically for the terms $\bb_1, 
E_2$, and $\bb_3$, if the symplectic basis (SB; red) is made use of as 
discussed in \SEC\ \ref{sec:NF-step1}.
}
\label{fig:res-gen}
\end{figure}

\begin{proof}
\EDITS{
The proof works as follows (see \FIG\ 
\ref{fig:res-gen}):
\begin{enumerate}[(i)]
\item 
It is assumed that \EQS\ \eqref{eq:NF-integration-conditions-b} and 
\eqref{eq:NF-integration-conditions-c} are fulfilled for $\abs{\mm}=n+1$, \ie\ 
the respective term of the dynamical equations can be written as a symplectic
gradient
\begin{equation}
 \bb_n(\xx) = \symplecticJ \pop{}{\xx} \tilde E_{n+1} (\xx) 
\end{equation}
with a scalar function $\tilde E_{n+1} = \sum_{\abs{\mm}=n+1} \tilde \xi_\mm \,
\xx^\mm$.
Note that the latter does not need to be identical to the energy functional,
$\tilde E_{n+1} \neq E_{n+1}$.
\item
Replacing $\beta_{\mm k} \to \tilde\beta_{\mm k}$ and $\xi_\mm \to 
\tilde\xi_\mm$, \EQ\ \eqref{eq:NF-integration-conditions-a} with 
$\abs{\mm}=n+1$ 
as well as \eqref{eq:NF-integration-conditions-c} with $\abs{\mm}=n+2$ are used 
as conditional equations for the determination of the resonant coefficients of 
the order $n$. These equations form a linear system of equations with the 
resonant coefficients being the variables.
\item
By construction of step (ii), the assumption in step (i) is fulfilled in the
next-higher order $n+2$. Therefore, the procedure can be repeated successively
for the orders $n=3,5,7,\ldots$
\end{enumerate}
After having calculated the term $\tilde E_{n+1}$ in step (i), the system of 
equations in step (ii) can be set up by inserting \EQS\ 
\eqref{eq:NF-res-trafo-E-gn-coeff} and \eqref{eq:NF-res-trafo-gn-b3-coeff} into 
the \EQS\ \eqref{eq:NF-integration-conditions-a} and 
\eqref{eq:NF-integration-conditions-c} after having replaced $\beta_{\mm k} \to 
\tilde\beta_{\mm k}$ and $\xi_\mm \to \tilde\xi_\mm$.
}
%
%

\EDITS{
Equation \eqref{eq:NF-res-LGS-a} must hold for all $\abs{\mm} = n+1$ and \EQ\
\eqref{eq:NF-res-LGS-b} for all $\abs{\mm} = n+2$ as well as 
$i,j=1,\ldots,\dof$ 
with $i \neq j$.
Equations \eqref{eq:NF-res-LGS} are written in a way that all terms which depend
on the resonant coefficients $\gamma_{\mm k}$ occur on the left-hand side, while
the right-hand side is independent of these terms.
Because of the fact that the resonant coefficients enter \EQS\
\eqref{eq:NF-res-LGS} linearly according to \EQS\ \eqref{eq:NF-res-def-sigmas},
they form a linear system of equations which can formally be written as
\begin{equation}
  \mathcal{A} \, \mathcal{G} = \mathcal{B} \,.
\label{eq:NF-res-LGS-short}
\end{equation}
Here, the matrix $\mathcal{A}$ and the vector $\mathcal{B}$ are both determined 
by the known quantities $\beta_{\mm k}$, $\lambda_i$, $\xi_\mm$, $\tilde 
\xi_\mm$, and $m_k$.
All the unknown terms, \ie\ the resonant coefficients $\gamma_{\mm k}$, are
collected in the vector $\mathcal{G}$.
The number of the resonant coefficients is, in general, 
smaller than the number of equations, so that the system 
\eqref{eq:NF-res-LGS-short} is overdetermined. However, it is guaranteed by 
Darboux's theorem that there exists a solution, because otherwise the latter 
would be violated and it would not be possible to construct canonical 
coordinates.
}
\end{proof}

We note that it is appropriate to solve \EQ\ \eqref{eq:NF-res-LGS-short} via a 
least-square fit $\| \mathcal{A} \mathcal{G} - \mathcal{B} \| \stackrel{!}{=} 
\text{min}$. The minimum of this fit must be zero because of Darboux's theorem, 
and its actual value in a numerical implementation is a measure of success of 
the procedure.
After the resonant coefficients of a certain degree $n$ have been determined as 
the solutions of \EQ\ \eqref{eq:NF-res-LGS-short}, the corresponding 
transformation needs to be applied to the full expansion, \ie\ \EQS\ 
\eqref{eq:NF-time-one-map-trafo-vector} and 
\eqref{eq:NF-time-one-map-trafo-scalar} must be evaluated.

\subsection{Canonical torus structure of the noncanonical Hamiltonian system}
\label{sec:NF-canonstruc}

After the normal form expansion has been applied as discussed in \SECS\
\ref{sec:NF-step2} and \ref{sec:NF-step3}, the normal form coordinates
are canonical ones by construction. The expansions fulfil the canonical
equations \eqref{eq:NF-canonequations-for-NF} in every order $n$ with the
energy functional acting as Hamiltonian
\begin{equation}
  H = E(\xx) \,. 
  \label{eq:NF-E-as-Ham}
\end{equation}

\begin{thm}[Final structure of the normal form Hamiltonian]
\label{last-theorem}
\EDITS{
In the final set of coordinates, the products $x_{2i-1} x_{2i}$ are constants 
of motion and the transformed energy functional is dierctly given in terms of 
action variables $\JJ$,
\begin{equation}
  H = E(\JJ) = \sum_{n=0}^m E_n (\JJ) \,.
  \label{eq:NF-E-as-Ham-J}
\end{equation}
%
}
\end{thm}

\begin{proof}
\EDITS{
Because of the summation over $\mm \in \Mset$ in \EQ\
\eqref{eq:NF-polynomial-structure-E}, this Hamiltonian only consists of
monomials
\begin{equation}
  \xx^{\mm} 
  \bigl. \bigr|_{\mm \in \Mset}
  =
  \prod_{i=1}^\dof (x_{2i-1} x_{2i})^{m_{2i}}
  \defined
  \prod_{i=1}^\dof (q_i \, p_i)^{m_{2i}} \,, 
\end{equation}
where the normal form coordinates can be interpreted pairwise as standard
canonical coordinates $q_i \defined x_{2i-1}$ and $p_i \defined x_{2i}$ in the
last step.
Moreover, these products directly define action variables
\begin{equation}
  J_i \defined
  \begin{cases}
    \phantom{\ui} q_i p_i \,,
    & \lambda_i = \kappa_i \phantom{\ui\omega_i}
    (\kappa_i \in \Rbbm) \,,
\\
    \ui  q_i p_i \,,
    & \lambda_i = \ui \omega_i \phantom{\kappa_i}
    ( \omega_i \in \Rbbm) \,,
  \end{cases} 
  \label{eq:NF-action-coord-def}
\end{equation}
where the imaginary unit in the definition compensates the respective 
contribution of each purely imaginary eigenvalue. 
If all the action variables correspond to stable oscillations of the system, and
denoting the corresponding angle variables by $\varphi_i$, the dynamical
equations are
\begin{subequations}
\begin{align}
  \dot\varphi_i   &= \phantom{-} \pop{H(\JJ)}{J_i} \defined \omega_i(\JJ) \,,
   \\
  \dot J_i   &= - \pop{H(\JJ)}{\varphi_i} = 0 \,, 
\end{align}%
\label{eq:NF-action-angle-Eqs}%
\end{subequations}
where $\omega_i(\JJ)$ are the characteristic frequencies of the system. These
have the solution
\begin{subequations}%
\begin{align}
   \varphi_i(t) &= \omega_i t + \varphi_i(0) \,,   \\
   J_i(t)    &= \text{const.}  
\end{align}%
\label{eq:NF-action-angle-Eqs-sol}%
\end{subequations}
}
\end{proof}

In normal form coordinates, the dynamics of the system is
restricted to a $\dof$-dimensional torus $\mathcal{T}^\dof$ if all eigenvalues
are purely imaginary. 
If $k$ of the actions \eqref{eq:NF-action-coord-def} correspond to real
eigenvalues, then the dynamics takes place on a manifold with the structure
\EDITS{$\mathcal{T}^{\dof-k} \times \Rbbm^k$.}
As already mentioned above, an important case in the field of reaction dynamics 
in variational space is that of rank-1 saddle points ($k=1$). These form 
transition states where the reaction channel is given by the unstable direction 
of the saddle. If the normal form Hamiltonian \eqref{eq:NF-E-as-Ham-J} has been 
constructed at such a point the corresponding quantum reaction rate is directly 
given by \EQS\ \eqref{eq:directional-flux} and \eqref{eq:thermal-rate}.

\EDITS{
We finally note that our procedure basically also applies if the original 
coordinates are canonical ones and in this case, the procedure presented here 
merges the treatment already described by Waalkens \etal \cite{Waalkens2008}.
Technically, this is expressed in the relation that, for canonical coordinates, 
the generating function $W$ in \EQ~\eqref{eq:NF-H-CNF} is a special case of the 
generator $\vec\generator$.
The precise relation between the respective generating functions is then
$\vec\generator(\xx) = \symplecticJ\times[\partial W(\xx)/\partial\xx]$
in which all resonant terms $\gamma_{\mm i}$ in \EQ~\eqref{eq:NF-res-LGS} are 
identically zero.
}

\subsection{Implementation}

The procedure to construct local normal form Hamiltonians in action-angle 
variables presented above has the advantage that it can step-by-step be 
implemented using symbolic computations. 
\EDITS{
In the appendix, we provide both an exemplary code in the script language 
\textsc{Mathematica} which allows for the constructive application of the 
symbolic computation scheme in arbitrary normal form orders $m$ and for 
arbitrary degrees of freedom $\dof$ as well as the intermediate numerical 
results of the procedure.
}
As input, the code uses the definitions of the matrix $K$, the vector $\hh$ 
in \EQS\ \eqref{eq:TDVP-K-h-real} and the energy functional 
\eqref{eq:TDVP-energy-functional}, as well as the values of the respective 
degrees of freedom and the desired normal form order.
The output is the normal form Hamiltonian \eqref{eq:NF-E-as-Ham-J} in 
action-angle coordinates. This general numerical implementation allows for 
practical applications of the procedure, such as \eg\ the calculation of thermal 
decay rates in Bose-Einstein condensates in Refs.\ \cite{Junginger2012d, 
Junginger2013b, Junginger2012a, Junginger2012b, Junginger2015a}.

\section{Conclusion and outlook}

In this paper, we have demonstrated a general method to construct local, 
canonical coordinates in the vicinity of a fixed point of a noncanonical 
Hamiltonian system via normal form expansions.
The method allows for the general and systematic investigation of \eg\ quantum 
systems which are described within a variational approach.
It is applicable for systems with arbitrary degrees of freedom, in arbitrary 
order of the local expansion, it is independent of the precise form of the 
Hamilton operator, and it can be implemented in symbolic computations step by 
step.

Extensions and generalizations of the procedure will be necessary in case of 
zero-eigenvalues [$\lambda_i=0$ in \EQ\ 
\eqref{eq:TDVP-EV-structure-pm-structure}], degenerate ones or in case of 
higher-order resonances $\left(\left<\mm,\vec\lambda^\pm\right> = \lambda_k^\pm 
\right)$.
This would allow one to construct the system's canonical normal form also in 
situations with strong mode coupling of the different degrees of freedom. 
Further reductions of the constructed normal form Hamiltonian can be performed 
\eg\ using the methods of hypernormal forms and spectral sequences 
\cite{Kokubu1996, Baider1989, Chen2000, Murdock2004, Sanders2003} which can also 
be used in case of zero-eigenvalues of the linearised system.

\section*{Acknowledgement}

This work was supported by Deutsche Forschungsgemeinschaft. A.\,J.\ is grateful 
for support from the Landesgraduiertenf\"orderung of the Land 
Baden-W\"urttemberg. We thank Dario Bambusi, Marcel Griesemer, Guido Schneider, 
and the members of the Graduiertenkolleg 1838 ``Spectral Theory and Dynamics of 
Quantum Systems'' for fruitful discussions.

\appendix
\section{Numerical example}
\label{app:NF-example}

\EDITS{
In this appendix, a numerical example of the normal form procedure discussed 
in the present paper is presented.
In the following, both an exemplary \textsc{Mathematica} script code as well 
as important intermediate results of the calculations are shown. 
}

\EDITS{
The system considered is a \BEC\ with additional $1/r$-interaction which has 
already been discussed in detail in \REF~\cite{Junginger2012b}, and it is 
described within a variational approach \eqref{eq:TDVP-trial-wave-function} 
consisting of $\Ng=2$ coupled Gaussian wave functions,
\begin{equation}
 \psi(\rr,t) = \sum_{k=1}^\Ng g_k(\rr,t)
 \quad \text{with} \quad
 g_k = \ue^{-a_k r^2 + \gamma_k} \,.
\end{equation}
%
A detailed description of the variational approach's application is also given 
in \REF~\cite{Junginger2012b}, so that we concentrate in the following on the 
normal form procedure.
Note further that we set the physical parameters in \REF~\cite{Junginger2012b} 
to $\gamma_\text{trap}^2=2.5\times 10^{-4}$ and $a=-0.99$.
}

\lstset{
  language=Mathematica,
  basicstyle=\small\ttfamily\color{blue},
  commentstyle=\color{gray},
  backgroundcolor=\color{gray!10},
  morecomment=[s][\color{gray}]{(*}{*)},
  mathescape=true,
  }
\newcommand{\mtext}[1]{\text{\texttt{#1}}}

\EDITS{
A \textsc{Mathematica} script can be set up as follows:
In a first step, the global parameters for the number $\Ng$ of coupled Gaussian 
wave functions, the number of degrees of freedom $\dof$, the physical 
parameters $a, \gamma_\text{trap}^2$ in the GPE as well as the maximum normal 
form order $n_\text{max}$ are set:
}
\begin{lstlisting}
  Ng     = 2;
  d      = 2Ng-1;
  a      = -0.99;
  $\gamma$trap2  = 2.5$\times$10$^{-4}$;
  nmax   = 7;
\end{lstlisting}
\EDITS{
Furthermore, the following commands are defined for the use in the subsequent 
script code:
}
\begin{lstlisting}
  x    = ToExpression["x" <> ToString[#]] &/@ Range[2d]; 
  y    = ToExpression["y" <> ToString[#]] &/@ Range[2d]; 
  m    = ToExpression["m" <> ToString[#]] &/@ Range[2d]; 
 
  ytox = ToExpression["y" <> ToString[#] <> "$\to$x" <> 
         ToString[#]] &/@ Range[2d]; 
  xtoy = ToExpression["x" <> ToString[#] <> "$\to$y" <> 
         ToString[#]] &/@ Range[2d]; 
  xoddtozero = ToExpression["x" <> ToString[2#-1] <> 
               "$\to$0"] &/@ Range[d]; 
  meventoodd = ToExpression["m" <> ToString[2#] <> "$\to$m" <> 
               ToString[2#-1]] &/@ Range[d]; 
  xtoj = ToExpression["x" <> ToString[2#-1] <> "$\to$j" <> 
         ToString[#] <> "/x" <> ToString[2#]] &/@ Range[d];
  z    = ToExpression["{a" <> ToString[#] <> ",$\gamma$" <> 
         ToString[#] <> "}"] &/@ Range[Ng]/.$\gamma$1$\to$0
  pwr[x_, m_] := Product[x[[i]]^m[[i]], {i,2d}];
\end{lstlisting}
%
\EDITS{
In order to evaluate the expectation values \eqref{eq:TDVP-energy-functional} 
of the single contributions of 
the Hamilton operator, it is appropriate to define the auxiliary functions 
}
\begin{lstlisting}
  i0[{a_,$\gamma$_}] := ($\pi$/a) Sqrt[$\pi$/a] Exp[-$\gamma$];
  i2[{a_,$\gamma$_}] := (3$\pi$/(2a^2)) Sqrt[$\pi$/a] Exp[-$\gamma$];
  im[{a_,$\gamma$_}] := $\pi$ Sqrt[$\pi$/a] Exp[-$\gamma$];
\end{lstlisting}
\EDITS{
with which the norm of the wave function and the normalized expectation values  
\eqref{eq:TDVP-energy-functional} can be evaluated as follows:
}
\begin{lstlisting}
  $\xi$[k_,l_] := z[[k]] + ComplexExpand[Conjugate[z[[l]]]];
  $\xi$[i_,j_,k_,l_] := z[[i]] + ComplexExpand[Conjugate[z[[j]]]] +
                    z[[k]] + ComplexExpand[Conjugate[z[[l]]]];

  N2    = Sum[ i0[$\xi$[k,l]] , {k,Ng}, {l,Ng}]; 
 
  Ekin  = Sum[ 6z[[k,1]] i0[$\xi$[k,l]] - 4z[[k,1]]^2 i2[$\xi$[k,l]]
             , {k,Ng}, {l,Ng}] / N2; 
      
  Etrap = $\gamma$trap2 Sum[ i2[$\xi$[k,l]] , {k,Ng}, {l,Ng}] / N2; 
 
  Ec    = 8$\pi$a  / N2^2
          Sum[ i0[$\xi$[i,j,k,l]], {i,Ng}, {j,Ng}, {k,Ng}, {l,Ng}];
     
  Emon  = -4$\pi$ Sum[ im[$\xi$[i,j,k,l]] / ($\xi$[i,j][[1]]  $\xi$[k,l][[1]])
                 , {i,Ng}, {j,Ng}, {k,Ng}, {l,Ng}] / N2^2; 
\end{lstlisting}
\EDITS{
Therewith, the energy functional and the vector $\hh$ in \EQ\ 
\eqref{eq:TDVP-h-real} can be obtained by
}
\begin{lstlisting}
  Emf = Ekin + Eext + (Ec + Emon)/2;
  h   = D[Emf, {Delete[Flatten[z], 2]}];
\end{lstlisting}
\EDITS{
With an appropriate set of initial values \texttt{z0}, a root search yields 
the fixed point of the dynamical equations, and the fixed point energy is 
obtained by inserting the parameters into the energy functional:
}
\begin{lstlisting}
  fp   = FindRoot[h, z0];
  emf0 = emf/.fp
\end{lstlisting}
\EDITS{
For the above given physical parameters, one fixed point is 
}
\begin{align} \small
  \mtext{a1} = \mtext{0.0631758} \,, \qquad 
  \mtext{a2} = \mtext{0.215522} \,, \qquad 
  \mtext{$\gamma$2} = \mtext{-0.481071} \,.
\end{align}
\EDITS{
The normalization of the wave function at this fixed point is explicitly taken 
into account in the script by dividing the expectation values by $\mtext{N2}$.
The fixed point corresponds to the ground state and it has an energy of 
}
\begin{equation} \small
  \mtext{emf0} = \mtext{-0.137356} \,.
\end{equation}
\EDITS{
After the fixed point has been determined it is shifted to the origin of the 
coordinate system:
}
\begin{lstlisting}
 yrel = ToExpression["y" <> ToString[2#-1] <> "+Iy" <> 
        ToString[2#]] &/@Range[d]; 
 yrel = Partition[Insert[yrel,0,2] ,2]; 
 z2   = z + $\epsilon$ yrel /.fp; 
 z1   = z2 /.ytox;
\end{lstlisting}
\EDITS{
Here, $\mtext{z1}$ and $\mtext{z2}$ are two independent sets of local 
coordinates at the fixed point.
In the next step, the TDVP\ is set up in these local coordinates: 
}
\begin{lstlisting}
 $\xi$[k_,l_] := z1[[k]] + ComplexExpand[Conjugate[z1[[l]]]];
 $\xi$[i_,j_,k_,l_] := z1[[i]] + ComplexExpand[Conjugate[z1[[j]]]] +
                   z1[[k]] + ComplexExpand[Conjugate[z1[[l]]]];
 
 N2    = Sum[ i0[$\xi$[k,l]] , {k,Ng}, {l,Ng}] // Simplify; 
    
 Ekin  = Sum[ 6z1[[k,1]] i0[$\xi$[k,l]] - 4z1[[k,1]]^2 i2[$\xi$[k,l]]
            , {k,Ng}, {l,Ng}] / N2; 
 
 Etrap = $\gamma$2 Sum[ i2[$\xi$[k,l]] , {k,Ng}, {l,Ng}] / N2; 
 
 Ec    = 8$\pi$a / N2^2
         Sum[ i0[$\xi$[i,j,k,l]] , {i,Ng}, {j,Ng}, {k,Ng}, {l,Ng}]; 
 
 Emon  = -4$\pi$ Sum[ im[$\xi$[i,j,k,l]] / ($\xi$[i,j] $\xi$[k,l])
                , {i,Ng}, {j,Ng}, {k,Ng}, {l,Ng}] / N2^2; 
\end{lstlisting}
\EDITS{
Therewith, the energy functional, the matrix $K$, as well as the vector $\hh$ 
can be defined in local coordinates at the fixed point, and they are expanded 
up to the order $\nmax$:
}
\begin{lstlisting}
  (* Matrix K *)
  S    = Simplify[Sum[ i0[(z1[[k]] + 
         ComplexExpand[Conjugate[z2[[l]]]])] , {k,Ng}, {l,Ng}]]; 
  Si   =  D[S, {x}] /.ytox; 
  Sj   =  D[S, {y}] /.ytox; 
  dnx  = -D[N2,{x}] / (2 Sqrt[N2]^3);  
  term = 2 Expand[(D[S,{x},{y}] /.ytox) / (N2 $\epsilon$^2) + 
         ( Outer[Times, Si/$\epsilon$, dnx/$\epsilon$] +
           Outer[Times, dnx/$\epsilon$, Sj/$\epsilon$] ) / Sqrt[N2]]; 

  kcoef      = Table[0, {nmax}]; 
  kcoef[[1]] = ComplexExpand[Im[Expand[term /.$\epsilon \to 0$]]]; 
 
  Do[ term       = D[term, $\epsilon$]/(n-1); 
      kcoef[[n]] = ComplexExpand[Im[Expand[term /.$\epsilon \to 0$]]]
    , {n, 2, nmax}]
  
  (* Vector h *)
  hcoef = Table[0, {nmax}]; 
  term  = D[Ekin + Etrap + (Ec + Emon)/2, $\epsilon$]; 
 
  Do[ term       = D[term, $\epsilon$]/(n+1); 
      hcoef[[n]] = ComplexExpand[Re[Expand[
                   -D[term /.$\epsilon \to 0$, {x}]]]]
    , {n, nmax}];
  
  (* Energy functional Emf *)
  emfcoef = Table[0, {nmax-2}];
  term    = D[Ekin + Etrap + (Ec + Emon)/2, $\epsilon$]; 
 
  Do[ term         = D[term, $\epsilon$]/(n+1); 
      emfcoef[[n]] = Expand[term /. $\epsilon \to 0$]
    , {n, nmax-2}];
  
  (* Increase precision *)
  kcoef   = SetPrecision[kcoef   // Chop, 50]; 
  hcoef   = SetPrecision[hcoef   // Chop, 50]; 
  emfcoef = SetPrecision[emfcoef // Chop, 50];
\end{lstlisting}
\EDITS{
Up to this point, the script is adapted to the special physical system of a 
\BEC\ with $1/r$-interaction. The following script code, however, is 
independent of the system which is investigated. Only the structure of some
do-loops has to be adapted if calculations are performed with $\dof\neq 3$ 
degrees of freedom.
}

\subsection{Diagonalization and symplectic basis}

\EDITS{
After the steps performed above, the local expansions of the matrix $K$, the 
vector $\hh$ and the energy functional $E$ are known, and the single terms of 
the expansion are orderwise stored in the quantities $\mtext{kcoef}$, 
$\mtext{hcoef}$, and $\mtext{emfcoef}$. Consequently, the transformations can 
be applied as described in \SECS\ \ref{sec:NF-step1} to \ref{sec:NF-step3}.
As explained in \SEC\ \ref{sec:NF-step1}, the first step of the transformations 
is to diagonalize the system with respect to its linearized part. 
Therefore, the latter's eigenvalues and -vectors are required which 
can be obtained the following way:
}
\begin{lstlisting}
  k0    = kcoef[[1]]; 
  k0inv = Inverse[k0]; 
  
  Do[ kcoef[[i]] = k0inv.kcoef[[i]] // Expand; 
      hcoef[[i]] = k0inv.hcoef[[i]] // Expand
    , {i,nmax}];

  ktmp = kcoef; 
  htmp = hcoef; 
  
  Do[
      Do[ ktmp[[i+j]] = ktmp[[i+j]] -
                        kcoef[[i]].ktmp[[1+j]] // Expand
        , {j,0,nmax-i}]; 

      Do[ htmp[[i+j]] = htmp[[i+j]] - 
                        kcoef[[i]].htmp[[1+j]] // Expand
        , {j,0,nmax-i}]; 
        
      kcoef = ktmp; 
      hcoef = htmp
     
    , {i,2,nmax}];

  (* Eigenvalues and -vectors *)
  Jac = D[hcoef[[1]], {x}];
  $\lambda$    = Eigenvalues[Jac] // Chop;
  T   = Eigenvectors[Jac] // Chop;
\end{lstlisting}
\EDITS{
For the above mentioned parameters, the eigenvalues are
\begin{align} \small
  \lambda = 
  \{
  \mtext{$\pm$1.87399i, $\pm$0.855197i, $\pm$0.182227i}
  \}
  \label{eq:app-NF-example-ev}
\end{align}
}
\EDITS{%
and they possess the structure \eqref{eq:TDVP-EV-structure-pm-structure}. The 
matrix $\mtext{T}$ contains the eigenvectors of the linearized dynamical 
equations, whose symplectic normalization 
\eqref{eq:NF-eigenvectors-normalization} can be carried out as follows:
}
\begin{lstlisting}
  tk0tt = T.k0.Transpose[T] // Chop;
  Do[ T[[i]] = T[[i]] / tk0tt[[i-1,i]] , {i,2,2d,2}]; 
  jmat = T.k0.Transpose[T] // Chop;
\end{lstlisting}
\EDITS{
Finally, the transformation of the dynamical equations and the energy 
functional to the symplectic basis is obtained by 
}
\begin{lstlisting}
  emfcoef = Collect[emfcoef /.xtoy, y]; 
  hcoef   = Collect[hcoef   /.xtoy, y];
  
  ttx     = Transpose[T].x; 
  ytottx  = ToExpression[ "y" <> ToString[#] <> "$\to$ttx[[" <> 
            ToString[#] <> "]]" & /@ Range[2 d]]; 
  
  emfcoef = emfcoef /. ytottx // Expand; 
  hcoef   = hcoef   /. ytottx // Expand; 
  ttinv   = Inverse[Transpose[T]]; 
  
  Do[ hcoef[[i]] = ttinv.hcoef[[i]] // Expand // Chop, {i,nmax}]; 
\end{lstlisting}
\EDITS{
After this transformation to the symplectic basis, the dynamical equations have 
the diagonal linear term
}
\begin{equation} \small
  \mtext{hcoef[[1]]} = 
  \begin{pmatrix}
    \mtext{+1.87399i\;x1}\\
    \mtext{-1.87399i\;x2}\\
    \mtext{+0.855197i\;x3}\\
    \mtext{-0.855197i\;x4}\\
    \mtext{+0.182227i\;x5}\\
    \mtext{-0.182227i\;x6}
  \end{pmatrix} \,,
\end{equation}
\EDITS{
whose entries are the eigenvalues \eqref{eq:app-NF-example-ev} and the 
quadratic order of the energy functional is 
}
\begin{equation} \small
  \mtext{emfcoef[[1]]} =
  \mtext{1.87399i\;x1\,x2 + 0.855197i\;x3\,x4 + 0.182227i\;x5\,x6} \,.
\end{equation}
\EDITS{
At this point, it is obvious that the coordinates $\mtext{x1,x2,x3,x4,x5,x6}$ 
are pairwise canonical up to this order of the expansion.
It is noted that the higher-order terms $\mtext{hcoef[[i]]}$ and 
$\mtext{hcoef[[i]]}$ with $\mtext{i>1}$ have not been simplified by this step. 
In general, they still contain all possible monomials, and because of the huge 
amount of terms, they are not shown here.
}

\subsection{Lie transforms for truncated expansions}

\EDITS{
As discussed in \SEC\ \ref{sec:NF-step2}, these higher-order terms are 
simplified via a normal form expansion, which is performed order by order 
($\mtext{n=1,2,3,...}$). 
Since the dynamical equations are on hand as truncated Taylor expansions 
\eqref{eq:NF-multivariate-expansion}, the number of applications of the Lie 
operator in \EQ\ \eqref{eq:NF-time-one-map-trafo-vector} and the 
right-multiplication operator in \EQ\ \eqref{eq:NF-time-one-map-trafo-scalar} 
can be limited.
In addition, it is appropriate to apply the operators to the different orders
of the expansions separately. For a generating function $\vec\generator_n$ of
degree $n$, the corresponding transformations
\eqref{eq:NF-time-one-map-trafo-vector} and
\eqref{eq:NF-time-one-map-trafo-scalar} then read
}
\begin{subequations}
\begin{align}
  \sum_{j=0}^\infty 
  \frac{1}{j!} \Lop{\vec\generator_n}{j} \aa (\xx)
=
  \sum_{k=1}^\nmax
  \sum_{j=0}^\infty 
  \frac{1}{j!} \Lop{\vec\generator_n}{j} \aa_k (\xx)
&\quad\longrightarrow\quad
  \sum_{k=1}^{\nmax}
  \sum_{j=0}^{\jmax}
  \frac{1}{j!} \Lop{\vec\generator_n}{j} \aa_k (\xx) \,,
\\
  \sum_{j=0}^\infty 
  \frac{1}{j!} \Dop{\vec\generator_n}{j} E (\xx)
=
  \sum_{k=0}^{\nmax+1}
  \sum_{j=0}^\infty 
  \frac{1}{j!} \Dop{\vec\generator_n}{j} E_k (\xx)
&\quad\longrightarrow\quad
  \sum_{k=0}^{\nmax+1}
  \sum_{j=0}^{\jmax}
  \frac{1}{j!} \Dop{\vec\generator_n}{j} E_k (\xx) \,,
\end{align}
\label{eq:NF-truncated-introduce-max-values}%
\end{subequations}
\EDITS{
where the limit of the summation over $j$ has been reset to $\jmax$ in
the respective last steps.
}

\EDITS{
Both the expressions $\Lop{\vec\generator_n}{j} \aa_k (\xx)$ and 
$\Dop{\vec\generator_n}{j} E_k (\xx)$ occurring on the right-hand side of \EQS\ 
\eqref{eq:NF-truncated-introduce-max-values} are of the order $k + j(n-1)$.
Thus, if one focuses only on the $l$-th order of the transformed field, there
will only contribute such terms for which $k + j(n-1) \stackrel{!}{=} l$.
Consequently, it is sufficient to apply the operators no more than
}
\begin{equation}
  \jmax = \left\lfloor \frac{l-k}{n-1} \right\rfloor 
  \label{eq:NF-jmax}
\end{equation}
\EDITS{
times, where $\lfloor \cdot \rfloor$ denotes the integer part of its argument.
}

\EDITS{
Moreover, it is numerically appropriate not to apply the operators
$\Lop{\vec\generator_n}{j}$ and $\Dop{\vec\generator_n}{j}$ several times and
to add the respective terms afterwards, as the formal transformations in \EQS\
\eqref{eq:NF-truncated-introduce-max-values} suggest, but to calculate the
transformation via the Horner-like scheme
}
\begin{subequations}
\begin{align} 
 \sum_{j=0}^{j_\text{max}} \tfrac{1}{j!} \Lop{\vec\generator_n}{j} \aa_k(\xx) 
  &= 
 \aa_k(\xx) + \Lop{\vec\generator_n}{}
 \biggl( \aa_k(\xx) + \tfrac{1}{2} \Lop{\vec\generator_n}{}
 \Bigl ( \aa_k(\xx) + \ldots
 \bigl ( \aa_k(\xx) + \tfrac{1}{j_\text{max}} \Lop{\vec\generator_n}{}
         \aa_k(\xx)
 \bigr )
 \Bigr )
 \biggr) \,,
\\
 \sum_{j=0}^{j_\text{max}} \tfrac{1}{j!} \Dop{\vec\generator_n}{j} E_k(\xx) 
  &= 
        E_k(\xx) + \Dop{\vec\generator_n}{}
 \biggl( E_k(\xx) + \tfrac{1}{2} \Dop{\vec\generator_n}{}
 \Bigl ( E_k(\xx) + \ldots
 \bigl ( E_k(\xx) + \tfrac{1}{j_\text{max}} \Dop{\vec\generator_n}{}
         E_k(\xx) 
 \bigr )
 \Bigr )
 \biggr) \,,
\end{align}%
\label{eq:NF-Horner-scheme}%
\end{subequations}
\EDITS{
where the maximum value of $\jmax$ is determined by \EQ\ \eqref{eq:NF-jmax}.
For each order, the generating function is first constructed according to \EQ\ 
\eqref{eq:NF-coefficients-generator-construction}. Second, for each generating 
function, the corresponding transformations 
\eqref{eq:NF-time-one-map-trafo-vector} and 
\eqref{eq:NF-time-one-map-trafo-scalar} are evaluated:
}
\begin{lstlisting}
  m1 = n - Sum[m[[i]], {i, 2, 2d}];

  Do[ (* Construction of the generating function *)
      g = Table[0, {2d}];
 
      Do[ term = pwr[x, m];
        
          Do[ nen = $\lambda$.m - $\lambda$[[i]];
   
              If[ Abs[nen] > 10^(-10) ,
                  g[[i]] = g[[i]] + 
                  Coefficient[hcoef[[n, i]], term]/nen term; ]
                  
            , {i,2d}]
        
        (* The range of the loop is adapted to d=3 *)
        , {m6, 0, n}
        , {m5, 0, n - m6}
        , {m4, 0, n - m6 - m5}
        , {m3, 0, n - m6 - m5 - m4}
        , {m2, 0, n - m6 - m5 - m4 - m3}];
 
    (* Transformation of the dynamical equations *)
    g1   = Transpose[D[g, {x}]] // Expand; 
    jmax = Floor[(nmax-1)/(n-1)]; 
    len  = nmax - (n-1)*jmax; 
    b    = hcoef; 
    lg   = Table[0, {nmax}];
 
    Do[ lg[[n;;n+len-1]] = (D[b[[1;;len]],{x}].g - 
                            b[[1;;len]].g1) / (jmax+1-j);
        b = lg + hcoef // Expand; 
        len = len + n - 1
      , {j, jmax}];
 
    (* Transformation of the energy functional *)
    If[jmax > 0, {hcoef = b; hcoef[[n]] = hcoef[[n]] // Chop}];
    jmax = Floor[(nmax-3)/(n-1)]; 
    len  = nmax - 2 - (n-1)*jmax; 
    b    = emfcoef; 
    lg   = Table[0, {nmax-2}];
    
    Do[ lg[[n;;n+len-1]] = (D[b[[1;;len]], {x}].g) / (jmax+1-j); 
        b = lg + emfcoef // Expand; 
        len = len + n - 1
      , {j, jmax}];
 
    If[jmax > 0, {emfcoef = b; emfcoef[[n]] 
                  = If[OddQ[n], Chop[emfcoef[[n]],10^(-8)],0]}];
 
  , {n, 2, nmax}];
\end{lstlisting}
\EDITS{
The normal form transformations in this step have removed all terms of the 
dynamical equations of even order and the odd-order terms of the energy 
functional, \ie
}
\begin{subequations}
\begin{gather} \small
  \mtext{hcoef[[2]]} = 
  \mtext{hcoef[[4]]} = 
  \mtext{hcoef[[6]]} = 0 \,, \\
  \mtext{emfcoef[[2]]} = 
  \mtext{emfcoef[[4]]} = 
  \mtext{emfcoef[[6]]} = 0 \,.
\end{gather}
\end{subequations}
\EDITS{
The first-order terms of the dynamical equations and the second-order terms of 
the energy functional have been left unchanged. The next-higher order 
corrections read
}
\begin{align}\small
  \mtext{hcoef[[3]]} = 
  \begin{pmatrix} 
    \mtext{   420.512\;x1}^2 \mtext{x2}
    \mtext{ + 68.8326\;x1\,x3\,x4}
    \mtext{ - 3.2175\;x1\,x5\,x6} \\
    \mtext{ - 420.512\;x1\,x2}^2
    \mtext{ - 68.8326\;x2\,x3\,x4}
    \mtext{ + 3.2175\;x2\,x5\,x6} \\
    \mtext{   68.8326\;x1\,x2\,x3}
    \mtext{ + 47.0488\;x3}^2\,\mtext{x4}
    \mtext{ - 0.972187\;x3\,x5\,x6}\\
    \mtext{ - 68.8326\;x1\,x2\,x4}
    \mtext{ - 47.0488\;x3\,x4}^2
    \mtext{ + 0.972187\;x4\,x5\,x6} \\
    \mtext{- 3.2175\;x1\,x2\,x5}
    \mtext{ - 0.972187\;x3\,x4\,x5}
    \mtext{ + 0.712752\;x5}^2\,\mtext{x6} \\
    \mtext{3.2175\;x1\,x2\,x6}
    \mtext{ + 0.972187\;x3\,x4\,x6}
    \mtext{ - 0.712752\;x5\,x6}^2
  \end{pmatrix}
  \label{eq:NF-example-hcoef3}
\end{align}
\EDITS{
in the dynamical equations and
}
\begin{align}
\begin{split}\small
  \mtext{emfcoef[[3]]} = \,&
  \mtext{- 394.668\;x1}^2\,\mtext{x2}^2\mtext{ + 348.835\;x1\,x2\,x3\,x4} \\
  &\mtext{- 9.43647\;x3}^2\,\mtext{x4}^2 \mtext{ - 25.7015\;x1\,x2\,x5\,x6} \\
  &\mtext{- 8.61906\;x3\,x4\,x5\,x6 - 0.148958\;x5}^2\,\mtext{x6}^2
\end{split}
\label{eq:NF-example-emfcoef3}
\end{align}
\EDITS{
in the energy functional, which both possess the polynomial structures 
\eqref{eq:NF-polynomial-structure-DGLs} and 
\eqref{eq:NF-polynomial-structure-E}. The same polynomial structure is also 
present in the higher-order terms, but they are not shown because of the large 
number of monomials.
}

\subsection{Determination of the resonant coefficients}

\EDITS{
Obviously, the third-order terms \eqref{eq:NF-example-hcoef3} fulfill the 
conditions of integrability \eqref{eq:NF-integration-conditions-b} and 
\eqref{eq:NF-integration-conditions-c}. However, they are not connected to the 
fourth-order term \eqref{eq:NF-example-emfcoef3} via \EQ\ 
\eqref{eq:NF-integration-conditions-a}.
In order to achieve the fulfillment of the canonical equation, the 
transformation using the resonant terms of the generating function is applied 
as discussed in \SEC~\ref{sec:NF-step3}.
As an alternative to its explicit evaluation, the system of equations 
\eqref{eq:NF-res-LGS-short} is here set up by evaluating the term 
$\Lop{\vec\generator_n}{}\bb_3$ instead of calculating each component 
\eqref{eq:NF-res-LGS} separately. For this purpose, the resonant terms in the 
generating function are labeled $c_i$ and they are treated as free parameters 
in the Lie transform.
}

\EDITS{
In order to determine the resonant coefficients of a generating function, the
linear system of equations \eqref{eq:NF-res-LGS-short} must be solved. As
already discussed in \SEC\ \ref{sec:NF-step3}, this system is overdetermined, 
but it is guaranteed by Darboux's theorem that a solution exists. 
However, because of numerical errors, one may be prevented from finding an
\emph{exact} solution of the equations. Therefore, an appropriate way to find
the resonant coefficients is to apply a least-square fit 
\begin{equation}
  \| \mathcal{A} \, \mathcal{G} - \mathcal{B} \|^2 \stackrel{!}{=} \text{min.}
  \label{eq:NF-least-square-fit}
\end{equation}
to the system of equations \eqref{eq:NF-res-LGS-short}.
It is emphasized that \EQ\ \eqref{eq:NF-least-square-fit} is \emph{not} an
approximation to the solution of the resonant coefficients, because its minimum
value must be (numerically) zero. 
The least-square fit is rather a suitable method to solve the overdetermined
system of equations.
}
\begin{lstlisting}
  hint = Table[0, {nmax}]; n = 1;
  hint[[n]] = Expand[Sum[Integrate[
              hcoef[[n,2i-1]] /.xoddtozero[[1;;i-1]],x[[2i]]]
              , {i, d}]];

  Do[ hint[[n]] = Expand[Sum[Integrate[ hcoef[[n,2i-1]] 
                  /.xoddtozero[[1;;i-1]],x[[2i]]] , {i,d}]]; 
                
      hdiff = hint[[n]] - emfcoef[[n]];
      nc    = d Binomial[d + (n-3)/2, d-1]; 
      noe1  = Binomial[d + (n-1)/2, d-1]; 
      
      (* Generating function with free parameters *)
      c     = ToExpression["c" <> ToString[#]] & /@ Range[nc]; 
      g = 0; cnt = 0; eqcnt = 0; n1h = (n+1)/2; 
      eq = Table[1, {noe1}]; 
      m1 = n1h - Sum[m[[i]], {i,3,2d,2}];
  
      Do[ eqcnt = eqcnt + 1; 
       
          Do[ If[ m[[i]] > 0, {cnt = cnt + 1; 
                  eq[[eqcnt]] = eq[[eqcnt]] - c[[cnt]]; 
                  term = pwr[x, m /. meventoodd]; 
                  g = g + c[[cnt]] Coefficient[hdiff, term] 
                  term/x[[i]]/$\lambda$[[i]] UnitVector[2d, i+1]}
                ]
            , {i,1,2d,2}]
        
        (* The range of the loop is adapted to d=3 *)
        , {m5, 0, n1h}
        , {m3, 0, n1h - m5}];
  
    (* Transformation induced by the generating function *)
    g1   = Transpose[D[g,{x}]]; 
    htmp = hcoef[[n+2]] + D[hcoef[[3]],{x}].g - hcoef[[3]].g1; 
    n3h  = (n+3)/2; 
    m1   = n3h - Sum[m[[i]], {i,3,2d,2}];
  
    Do[ term = pwr[x, m /. meventoodd]; 
        
        Do[ If[m[[i]] > 0, 
  
            Do[ If[ m[[j]] > 0, {
                eqcnt = eqcnt + 1; AppendTo[eq, Expand[ 
                Coefficient[htmp[[i]], term/x[[i+1]]]/m[[i]] - 
                Coefficient[htmp[[j]], term/x[[j+1]]]/m[[j]]]]}]
              , {j,i+2,2d,2}] ]
          , {i,1,2d-2,2}]
          
      , {m5, 0, n3h}
      , {m3, 0, n3h - m5}];
  
    (* Determination the resonant coefficients *)
    cmat = D[eq, {c}]; 
    b    = -eq /.Table[c[[i]] $\to$ 0, {i,nc}]; 
    c    = LeastSquares[cmat, b];
    crep = ToExpression[ "c" <> ToString[#] <> "$\to$c[[" <> 
           ToString[#] <> "]]" & /@ Range[nc]];
    g    = g /. crep; 
    g1   = Transpose[D[g, {x}]];
    
    (* Transformation of the dynamical equations *)
    jmax = Floor[(nmax-1)/(n-1)]; 
    len  = nmax - (n-1)*jmax; 
    b    = hcoef; 
    lg   = Table[0, {nmax}]; 
    
    Do[ lg[[n;;n+len-1]] = (D[b[[1;;len]],{x}].g - 
                            b[[1;;len]].g1)/(jmax+1-j); 
        b = lg + hcoef // Expand; 
        len = len + n - 1
      , {j, jmax}]; 
    
    If[jmax > 0, hcoef = b];
    
    (* Transformation of the energy functional *)
    jmax = Floor[(nmax-3)/(n-1)]; 
    len  = nmax - 2 - (n-1)*jmax; 
    b    = emfcoef; 
    lg   = Table[0, {nmax-2}]; 
    
    Do[ lg[[n;;n+len-1]] = (D[b[[1;;len]],{x}].g)/(jmax+1-j); 
        b   = lg + emfcoef // Expand; 
        len = len + n - 1
      , {j, jmax}]; 
    
    If[jmax > 0, emfcoef = b]
    
  , {n,3,nmax-2,2}];
\end{lstlisting}
\EDITS{
The resonant coefficients of the generating function do only need to be 
determined up to the order $\nmax-2$. This guarantees the fulfillment of the 
conditions of integrability in the order $\nmax-1$, so that the energy 
functional in order $\nmax$ is obtained by a simple integration of the 
dynamical equations:
}
\begin{lstlisting}
  n = nmax;
  hint[[n]] = Expand[Sum[Integrate[
              hcoef[[n,2i-1]] /.xoddtozero[[1;;i-1]], 
              x[[2i]]], {i,d}]]; 

  (* Definition of action valiables *)
  jfac = Table[1, {nd}];
  jvar = Table[i, {nd}];
  Do[ If[Chop[Abs[Im[$\lambda$[[2i-1]]]]] > 10^-10,
         {jfac[[i]] = I, jvar[[i]] = j}]
    , {i, 1, nd}]
  actionvar= ToExpression[ "j" <> ToString[#] <> "$\to$" <> 
             ToString[jvar[[#]]] <> ToString[#] <> "/" <> 
             ToString[jfac[[#]]]] & /@ Range[nd]
    
  H = hint /.xtoj /.actionvar;
\end{lstlisting}
\EDITS{
The last step takes into account the definition of the action variables 
\eqref{eq:NF-action-coord-def} and it guarantees that the integrated 
Hamiltonian 
is real. In order to keep the information, which variables correspond to real 
(unstable) and imaginary (stable) eigenvalues, the coordinates are labeled 
"$\mtext{i}$" in the former and "$\mtext{j}$" in the latter case.
}

\EDITS{
Finally, the canonical equations are fulfilled in every order of the expansion 
by construction, and the local Hamiltonian in action coordinates orderwise 
consists of the terms
}
\begin{subequations}
\begin{align} \small
  \mtext{H[[1]]} = \;&
  \mtext{1.87399\;j1 + 0.855197\;j2 + 0.182227\;j3} \,,
\\[.5em]
\begin{split} \small
  \mtext{H[[3]]} =\;& 
  \mtext{- 210.256\;j1}^2\mtext{ - 68.8326\;j1\,j2 - 23.5244\;j2}^2 \\
  &\mtext{+ 3.2175\;j1\,j3 + 0.972187\;j2\,j3 - 0.356376\;j3}^2 \,,
\end{split}
\\[.5em]
\begin{split} \small
  \mtext{H[[5]]} =\;& 
  \mtext{37771.3\;j1}^3\mtext{ + 20967.8\;j1}^2\,\mtext{j2 - 10084.8\;j1\,j2}^2 
\\
  &\mtext{+ 861.764\;j2}^3\mtext{ - 1195.23\;j1}^2\,\mtext{j3 
  - 1505.82\;j1\,j2\,j3} \\
  &\mtext{- 104.915\;j2}^2\,\mtext{j3 + 118.681\;j1\,j3}^2
    \mtext{ + 48.8141\;j2\,j3}^2 \\
  &\mtext{- 2.74901\;j3}^3 \,,
\end{split}
\\[.5em]
\begin{split} \small
  \mtext{H[[7]]} = \;& 
  \mtext{- 9.61439$\times$10}^6\;\mtext{j1}^4\mtext{ 
  + 3.27764$\times$10}^7\;\mtext{j1}^3\,
  \mtext{j2 - 2.97357$\times$10}^7\;\mtext{j1}^2\,\mtext{j2}^2 \\
  &
  \mtext{+ 3.79731$\times$10}^6\mtext{\;j1\,j2}^3\mtext{ - 
  138082.\;j2}^4\mtext{ 
  + 315564.\;j1}^3\,\mtext{j3} \\
  &\mtext{+ 3.49666$\times$10}^6\;\mtext{j1}^2\,\mtext{j2\,j3}
  \mtext{ - 1.45983$\times$10}^6\;\mtext{j1\,j2}^2\,\mtext{j3}\\
  & \mtext{+ 39551.6\;j2}^3\,\mtext{j3} \mtext{ - 56668.5\;j1}^2\,\mtext{j3}^2 
  \mtext{ - 70531.1\;j1\,j2\,j3}^2 
  \\
  &\mtext{- 1471.93\;j2}^2\,\mtext{j3}^2 \mtext{ + 3042.69\;j1\,j3}^3 
  \mtext{ + 1240.61\;j2\,j3}^3 \mtext{ - 62.99\;j3}^4 \,.
\end{split}
\end{align}
\end{subequations}

\section*{Literature}

\end{document}